\documentclass[12pt]{article}

\usepackage[fleqn]{amsmath}
\usepackage{amsfonts}
\usepackage{amssymb}
\usepackage{ctable}
\usepackage[all,cmtip]{xy}
\usepackage{graphicx}
\usepackage{float}

\usepackage{subfigure}
\usepackage[title]{appendix}
\usepackage{xcolor}
\usepackage{hyperref}
\usepackage{bm}
\usepackage[margin=1.2in]{geometry}

\newcommand{\re}{^{\text{ret}}} 

\newcommand{\lt}[1]{\left#1}
\newcommand{\rt}[1]{\right#1}
\newcommand{\pt}[1]{\partial_{#1}}
\newcommand{\bsym}[1]{\boldsymbol{#1}}

\newcommand{\nh}{{\textbf{n}}}

\newcommand{\req}[1]{(\ref{eq:#1})}

\newcommand{\bld}[1]{\textbf{#1}}

\newcommand{\drm}{\mathrm{d}}

\newcommand{\Ddt}{\frac{\drm\phantom{s}}{\drm t}}

\newcommand{\pddt}{\frac{\partial\phantom{t}}{\partial t}}

\newcommand{\tpddt}{{\textstyle{\frac{\partial\phantom{t}}{\partial t}}}}

\newcommand{\refeq}[1]{(\ref{#1})}

\newcommand{\mbare}{m_{\text{b}}}
\newcommand{\mEL}{m_{\text{e}}}
\newcommand{\vect}[1] {\boldsymbol{{ #1}} }

\newcommand{\Rset}{\mathbb{R}}

\newcommand{\ID}{{\boldsymbol{I}_{3\times3}^{}}} % unit tensor

\newcommand{\aV}{\vect{a}}              % 3-acceleration

\newcommand{\fV}{\vect{f}}              % 3-force density vector field
\newcommand{\FV}{\vect{F}}              % 3-force vector
\newcommand{\jV}{{\vect{j}}}		% 3-current density
\newcommand{\nV}{{\vect{n}}}		% 3-unit vector
\newcommand{\qV}{{\vect{q}}}            % 3-position of particle
\newcommand{\sV}{{\vect{s}}}            % position 3-vector
\newcommand{\vV}{{\vect{v}}}            % 3-velocity of particle

\newcommand{\NullV}{\vect{0}}
\newcommand{\AV}{\pmb{{\cal A}}}
\newcommand{\BV}{\pmb{{\cal B}}}

\newcommand{\DV}{\pmb{{\cal D}}}
\newcommand{\EV}{\pmb{{\cal E}}}

\newcommand{\HV}{\pmb{{\cal H}}}

\newcommand{\PiV}{\boldsymbol{\Pi}}
\newcommand{\nab}{\vect{\nabla}}

\renewcommand{\leq}{\leqslant}
\renewcommand{\geq}{\geqslant}

\newcommand{\crprd}{{\boldsymbol\times}}

\newcommand{\DLWr}{{\DV}_{\text{\textsc{lw}}}^{\text{\tiny{ret}}}} %Lienard-Wiechert B field retarded
\newcommand{\HLWr}{{\HV}_{\text{\textsc{lw}}}^{\text{\tiny{ret}}}} %Lienard-Wiechert E field retarded

\newcommand{\ee}{\mathit{e}}

\newcommand{\eEL}{\mathit{e}}

\newcommand{\be}{\boldsymbol{\hat{e}}}

\newcommand{\bxi}{\boldsymbol{\xi}}

\newcommand{\bxio}{\boldsymbol{\xi}^\circ}

\newcommand{\bbeta}{{\boldsymbol{\beta}}}
\reversemarginpar

\numberwithin{equation}{section}

\begin{document}

%%%%%%%%%%%%%%%%%%%%%%%%%%%%%%%%%%%%%%%%%%%%%%%%%%%%%%%%%%%%%%%%%%%%%
%%%%%%%%%%%%%%%%%%%%%%%%%%%%%%%%%%%%%%%%%%%%%%%%%%%%%%%%%%%%%%%%%%%%%
%       
\title{\sc{radiation-reaction on the straight-line motion of a point charge accelerated by 
a constant applied electric field in an electromagnetic
 \sc{b}{\tiny\sc{opp}}-\sc{l}{\tiny\sc{and\'e}}-\sc{t}{\tiny\sc{homas}}-\sc{p}{\tiny\sc{odolsky}} vacuum}\vspace{-0.5truecm}}
\author{\textbf{Ryan J. McGuigan$^1$ and Michael K.-H. Kiessling$^2$}\\
\small           $^1$ Department of Physics, \\
\small           Lancaster University, \\
\small           Lancaster, \textcolor{black}{LA1 4YB,}  UK\\
\small           $^2$ Department of Mathematics, \\ 
\small           Rutgers, The State University of New Jersey,\\
\small           110 Frelinghuysen Rd., Piscataway, NJ 08854, USA\\ 
\textrm{\small Version of July 16, 2025. Typeset with \LaTeX\ on: }\vspace{-0.5truecm}}

\maketitle

\thispagestyle{empty}
\vspace{-1.1truecm}
\begin{abstract}\vspace{-.2truecm}
\noindent 
{The radiation-reaction problem of
standard Lorentz electrodynamics with point charges is pathological, standing in contrast to
Bopp--Land\'e--Thomas--Podolsky (BLTP) electrodynamics where it is in fact
well-defined and calculable, as reported in \cite{KiePRD}. 
To demonstrate the viability of BLTP electrodynamics, we consider the BLTP analogue of the radiation
reaction of a classical point charge accelerated from rest by a static homogeneous
capacitor plate field, and calculate it up to $O(\varkappa^4)$ in a formal expansion about $\varkappa=0$ 
in powers of $\varkappa$, Bopp's reciprocal length, a new electrodynamics parameter introduced by BLTP theory.
 In \cite{CarKie} the radiation-reaction corrections to test-particle motion were explicitly computed to 
$O(\varkappa^3)$, the first non-vanishing order.}
 In this article a crucial question regarding this ``small-$\varkappa$'' expansion, raised in \cite{CarKie}, is answered as 
follows: The motions computed with terms $O(\varkappa^3)$ included are mathematically accurate approximations to
{physically reasonable} solutions of the actual BLTP initial value problem for short times $t$, viz. 
when $\varkappa c t \ll 1$, where $c$ is the speed of light in vacuo, but their unphysical behavior over 
{much} longer times does not accurately approximate 
the actual BLTP solutions even when the dimensionless parameter $\varkappa e^2 / |\mbare| c^2 \ll 1$, 
where $e$ is the elementary charge and $\mbare$ the bare rest mass of the electron.
 This has the important implication that BLTP electrodynamics remains a viable contender for an accurate classical
electrodynamics with point charges that does not suffer from the infinite self-interaction problems of 
textbook Lorentz electrodynamics with point charges.\vspace{-0.2truecm}
\end{abstract}

\vfill
\hrule
\smallskip

\copyright(2026) \small{The authors. Reproduction of this preprint, in its entirety, is permitted

\hspace{1.5truecm} for non-commercial purposes only.}

\newpage

%%%%%%%%%%%%%%%%%%%%%%%%%%%%%%%%%%%%%%%%%%%%%%%%%%%%%%%%%%%%%%%%%%%%%%%%%%%

%%%%%%%%%%%%%%%%%%%%%%%%%%%%%%%%%%%%%%%%%%%%%%%%%%%%%%%%%%%%%%
%%%%%%%%%%%%%%%%%%%%%%%%%%%%%%%%%%%%%%%%%%%%%%%%%%%%%%%%%%%%%%
%%%%%%%%%%%%%%%%%%%%%%%%%%%%%%%%%%%%%%%%%%%%%%%%%%%%%%%%%%%%%%%%%%%%
%%%%%%%%%%%%%%%%%%%%%%%%%%%%%%%%%%%%%%%%%%%%%%%%%%%%%%%%%%%%%%
                \section{Introduction} \vspace{-10pt}
%%%%%%%%%%%%%%%%%%%%%%%%%%%%%%%%%%%%%%%%%%%%%%%%%%%%%%%%%%%%%%%%%%%%
%%%%%%%%%%%%%%%%%%%%%%%%%%%%%%%%%%%%%%%%%%%%%%%%%%%%%%%%%%%%%%
%%%%%%%%%%%%%%%%%%%%%%%%%%%%%%%%%%%%%%%%%%%%%%%%%%%%%%%%%%%%%%
%%%%%%%%%%%%%%%%%%%%%%%%%%%%%%%%%%%%%%%%%%%%%%%%%%%%%%%%%%%%%% 
\noindent
It has long been a well established fact that both the classical Lorentz electrodynamics 
\cite{Spohn} and Quantum Electrodynamics (QED) \cite{QEDbook} are plagued by infinities due
to the ``self''-interactions\footnote{The scare quotes around ``self''
 are meant to remind the reader that only the total electromagnetic fields are well-defined in principle, while
their splitting into ``self'' and ``non-self'' fields is to some extent ambiguous.
 Having stated this, from hence on we will omit those scare quotes.}
of point charges.
As far back as the 1930s, Max Born argued \cite{BornA, BornB} that some of these may disappear if, 
before quantization, one first fixes the infinite field-energy problem of classical Lorentz electrodynamics with point charges 
by modifying the classical law of the electromagnetic vacuum, formerly known as Maxwell's ``law of the pure ether''
(viz. $\HV=\BV$ and $\EV=\DV$).
 Joined in his quest soon after by Infeld \cite{BornInfeldA,BornInfeldB,BornInfeldCa,BornInfeldCb}, 
the (what we call) ``Born--Infeld (BI) law of the electromagnetic vacuum'' was proposed and subsequently 
found to be distinguished among all local and non-differential vacuum laws by being the only one that 
produces a field theory which, at least in principle, satisfies a handful of quite compelling postulates, thus: 
(1) it derives from an action principle that involves only the invariants 
$|\EV|^2-|\BV|^2$ and $\EV\cdot\BV$ in some algebraic or even transcendental way, and invariant $\rho\phi-\jV\cdot\AV$
linearly; hence is 
(1a) Lorentz-invariant and 
(1b) gauge-invariant (on account of the continuity equation being satisfied jointly by $\rho$ and $\jV$); 
(2) in the weak-field limit it reduces to the Maxwell--Lorentz field theory;
(3) the field energy density of the solutions is locally integrable, in particular also in the presence of 
moving point charge sources; 
(4) the solutions to the field equations do not exhibit birefringence.
 Postulates (1) to (3) are manifestly compelling, while (4) does hold for the linear Maxwell law of the pure ether and,
plausibly, also for the nonlinear regime. 

 Whether all these postulates are rigorously satisfied by the Maxwell--Born--Infeld field equations is still not completely 
settled.
 While it has been shown that electrostatic solutions of the Maxwell--Born--Infeld field equations
 with finitely many point charge sources and vanishing field conditions at spatial infinity do exist and have finite
field energy \cite{KieCMP}, it is still not rigorously known whether time-dependent solutions of these field equations 
do have finite field energy and momentum when the point charge sources are accelerated.
 In this vein, even the Maxwell--Born--Infeld field theory with prescribed reasonably moving point charge sources 
is not yet established as being mathematically well-posed, though conceivably and hopefully it will be in the not 
too distant future.

 The difficult nonlinearity of the BI vacuum law led Fritz Bopp \cite{BoppA,BoppB} to suggest as a linear 
alternative a vacuum law that involves higher order derivatives, viz. $\HV=(1+\varkappa^{-2}\square)\BV$ and 
$(1+\varkappa^{-2}\square)\EV=\DV$; here, $\varkappa$ is Bopp's reciprocal length parameter, and 
$\square:= \frac{1}{c^2}\partial_t^2-\Delta$ is the classical wave operator.
  Subsequently and independently the same vacuum law was proposed also by Land\'e and Thomas \cite{Lande,LandeThomas}, 
and picked up on by Podolsky \cite{Podolsky}.
 This model is often referred to by a subset of these names; to honor all initial contributors we 
will speak of the ``BLTP law of the electromagnetic vacuum.'' 

 Incidentally, Bopp's paper \cite{BoppA} seems to have inspired {Feynman's \cite{Feynman} relativistic UV 
cut-off for QED}; higher-order spin-offs followed, see \cite{PauliVillars} 
{for QED, and \cite{LazarLeckA} for classical electrodynamics.}
 Since the UV cutoffs in QED are non-pertubatively 
thought to be irremovable, and even when removed perturbatively do not lead to
a convergent perturbative expansion \cite{Dyson}, one may as well investigate what Feynman's UV cutoff implies
in the classical limit of QED and study BLTP electrodynamics.

However, the (putative) absence of the infinite self-field energies of point charges in a BLTP 
vacuum\footnote{Incidentally, we note that point dipoles still have infinite self-field energy in BI-  \cite{Erwin} and
in BLTP-electrodynamics \cite{LazarLeckB}.}
 only takes care of part of the problem.\footnote{Neither Born and Infeld, nor Bopp, nor Land\'e and Thomas, nor Podolsky 
gave much thought to the ill-defined self-force problem.
 Their proposals all involved the Lorentz formula for the electromagnetic force on a point charge in one way or another,
and thus they did not produce a well-defined electrodynamics. 
 Perennial claims to the contrary (cf. \cite{QuinnWald}, \cite{GHW}, \cite{PoissonETal}) notwithstanding, 
a generally acceptable way out of the dilemma did not materialize by elevating to an ``axiom'' that the electromagnetic
force on a point charge would have to be defined by averaging the Lorentz force field over some neighborhood of the
point charge's location \cite{Dirac}; see \cite{KiePRD} for a detailed critique.}

 A well-defined electromagnetic force on point charges was recently formulated by adapting Poincar\'e's definition of the 
electrodynamic force on a single extended charged particle in a Maxwell vacuum to many
point charges in a BI or BLTP vacuum \cite{KiePRD}. 
 In \cite{KiePRD} it was also reported that the joint dynamical problem for fields and particles 
is well-posed in {this formulation of BLTP electrodynamics.}
 A proof will appear in \cite{KTZonBLTP}.

 Also the scattering problem for a single particle that encounters a localized potential is well-posed \cite{VuMaria}
in BLTP electrodynamics.
 These authors showed that in this problem the self-force formula of \cite{KiePRD} can be converted into a formal 
Lorentz-type expression that involves integration over the whole past of the particle motion, first proposed in
\cite{LandeThomas} and further studied in \cite{Zayats} and \cite{GratusETal}.
 
 The important next step is to study the solutions of the joint initial value problem in BLTP electrodynamics, but
this is not straightforward.
 The simplest dynamical problem is the motion of a point charge along a constant applied electric field. 
 This is also a litmus test that well-known older models (in particular, 
the Landau--Lifshitz and Eliezer--Ford--O'Connell equations of motion) fail --- by yielding no radiation-reaction at all in 
this particular setting; cf. \cite{PMD}, \cite{KiePRD}, and \cite{CarKie} for detailed discussions.
 In \cite{CarKie} it was shown that BLTP electrodynamics passes this litmus test; i.e., 
radiation-reaction effects do in fact show up also in this simplest situation.
 More precisely, they show up at $O(\varkappa^3)$ when the electromagnetic force is formally expanded in powers of $\varkappa$
and then truncated by purging all terms proportional to $\varkappa^k$ with $k>K$ natural numbers; in this sense, 
radiation-reaction effects show up when $K\geq 3$, though not when $K\leq 2$.

 Interestingly, the solution to the equations of motion truncated after order $\varkappa^3$ can be computed 
in closed form, see \cite{CarKie}.
 Its long-time features are quite unexpected.
 When the bare rest mass of the electron $\mbare >0$,
 the solution looks physically reasonable for short times, following closely the radiation-free test 
particle motion but lagging behind a little due to radiation-damping --- as anticipated; yet over longer times the 
motion is periodic and does not resemble anything seen in linear accelerators.
 For $\mbare < 0$, the $O(\varkappa^3)$ solution of \cite{CarKie} resembles the radiation-free test 
particle motion over arbitrarily long times, but these negative $\mbare$ motions are in the opposite direction 
of what is seen in linear accelerators. 

 It is therefore of importance to answer the question whether ``small-$\varkappa$ regime'' means that 
the $O(\varkappa^3)$ approximation is accurate only for short times, viz. $\varkappa ct\ll 1$
where $c$ is the speed of light in vacuo, or whether it is accurate also when $\varkappa ct\gg 1$
provided the dimensionless parameter ${\varkappa e^2 }/{ |\mbare| c^2} \ll 1$, where $e$ is the elementary charge.
 If the $O(\varkappa^3)$ approximation is faithful even for $\varkappa ct\gg 1$ whenever ${\varkappa e^2 }/{ |\mbare| c^2}\ll 1$, 
then BLTP electrodynamics may have to be eliminated for good from the list of contenders for a realistic classical 
electrodynamics with point charges.
 A final verdict would require ${\varkappa e^2 }/{ \mbare c^2} \approx -2$ \cite{CKP}, though.

 In this paper radiation-reaction effects on the motion are computed {to order $\varkappa^4$ included.}
 The equations of motion truncated after the order-$\varkappa^4$ terms do not anymore appear 
to be solvable in closed form, but they can easily be solved numerically with the help of a computer, {and 
their solutions used to judge the accuracy of the $O(\varkappa^3)$ motions computed in \cite{CarKie}.}

 In the following, for the convenience of the reader we first summarize the dynamical equations and their
reduction to a Volterra equation, from \cite{CarKie}.
 Next we state the self-force formula evaluated to terms of order $\varkappa^4$ included, then we present
the numerically computed solutions for a representative selection of parameter values, and we end 
with a summary {of our results and an} outlook on future work.
 Many technical details of the expansion of the self-force at order $\varkappa^4$ are supplied in
Appendix A and B, for the perusal of the reader.

%%%%%%%%%%%%%%%%%%%%%%%%%%%%%%%%%%%%%%%%%%%%%%%%%%%%%%%%%%%%%%
%%%%%%%%%%%%%%%%%%%%%%%%%%%%%%%%%%%%%%%%%%%%%%%%%%%%%%%%%%%%%%
%%%%%%%%%%%%%%%%%%%%%%%%%%%%%%%%%%%%%%%%%%%%%%%%%%%%%%%%%%%%%%%%%%%%
%%%%%%%%%%%%%%%%%%%%%%%%%%%%%%%%%%%%%%%%%%%%%%%%%%%%%%%%%%%%%%
     \section{\hspace{-10pt}Initial value problem for fields and point charge} %\vspace{-10pt}
%%%%%%%%%%%%%%%%%%%%%%%%%%%%%%%%%%%%%%%%%%%%%%%%%%%%%%%%%%%%%%%%%%%%
%%%%%%%%%%%%%%%%%%%%%%%%%%%%%%%%%%%%%%%%%%%%%%%%%%%%%%%%%%%%%%
%%%%%%%%%%%%%%%%%%%%%%%%%%%%%%%%%%%%%%%%%%%%%%%%%%%%%%%%%%%%%%
%%%%%%%%%%%%%%%%%%%%%%%%%%%%%%%%%%%%%%%%%%%%%%%%%%%%%%%%%%%%%% 

 The electromagnetic vacuum in BLTP electrodynamics is defined by the two equations
\begin{alignat}{1}
        \HV(t,\sV)  
& = \label{eq:BLTPlawBandH}
       \left(1  + \varkappa^{-2}\square\,\right) \BV(t,\sV) \, , \\ 
        \DV(t,\sV) 
& =
        \left(1  + \varkappa^{-2}\square\,\right) \EV(t,\sV) \, ;
\label{eq:BLTPlawEandD}
\end{alignat}
in \refeq{eq:BLTPlawBandH} and \refeq{eq:BLTPlawEandD}, the parameter $\varkappa^{-1}$ is the ``Bopp length'' \cite{BoppA,BoppB},
and $\square \equiv c^{-2}\partial_t^2 -\Delta$ is the d'Alembertian, with $c$ the vacuum speed of light.
 The evaluations $\HV(t,\sV)$, $\BV(t,\sV)$, $\EV(t,\sV)$, and $\DV(t,\sV)$ of the fields at the space point
$\sV\in\Rset^3$ and instant of time $t\in\Rset$ are defined in any convenient flat foliation of Minkowski spacetime 
into space \&\ time. 
 These fields satisfy the familiar system of pre-metric Maxwell field equations, which consist of two evolution equations
\begin{alignat}{1}
\textstyle
\pddt{\BV(t,\sV)}
&= \label{eq:MdotB}
        - c \nab\crprd\EV(t,\sV) \, ,
\\
\textstyle
\pddt{\DV(t,\sV)}
&= 
        + c\nab\crprd\HV(t,\sV)  - 4\pi \ee \delta_{\qV(t)}(\sV){\vV}(t)\, ,
\label{eq:MdotD}
\end{alignat}
and two constraint equations
\begin{alignat}{1}
        \nab\cdot \BV(t,\sV)  
&= \label{eq:MdivB}
        0\, ,
\\
        \nab\cdot\DV(t,\sV)  
&=
        4 \pi \ee \delta_{\qV(t)}(\sV)\, .
\label{eq:MdivD}
\end{alignat}
 Here,  $\ee (>0)$ is the elementary electric charge,
$\qV(t)\in\Rset^3$ its position and $\vV(t) \in\Rset^3$ its velocity at time $t$.

  The particle's velocity is defined as usual to be the time-derivative of its position vector,
\begin{equation}
\Ddt \qV(t)
= : \label{eq:dotQisV}
\vV(t).
\end{equation}
   In the relativistic generalization of Newton's point mechanics by Einstein, Lorentz, and Poincar\'e, 
the velocity $\vV(t)$, in turn, changes with time according to
\begin{equation}
\Ddt   \frac{\vV(t)}{\sqrt{1 - \frac{1}{c^2}|{\vV}(t)|^2}}
= \label{eq:EinsteinNewtonEQofMOT}
\frac{1}{\mbare}\fV(t);
\end{equation}
here, $\mbare \neq 0$ is the \emph{bare inertial rest mass} of the particle, 
and $\fV(t)$ is the total electromagnetic force acting on it.
 Following Poincar\'e (cf. \cite{MillerBOOK}) we~define~it~as  (cf. \cite{KiePRD})
\begin{equation}
\fV(t) := \label{eq:POINCAREforce}
\eEL\,\EV^{\mbox{\tiny{hom}}}
 - \Ddt \int_{\Rset^3} \PiV^{\mbox{\tiny{field}}}(t,\sV) \drm^3s,
\end{equation}
where  $\EV^{\mbox{\tiny{hom}}}$ is a constant applied electric field (an idealization of the field between the plates of a 
capacitor), and $\PiV^{\mbox{\tiny{field}}}(t,\sV)$ is the momentum vector-density of the Maxwell-BLTP fields 
\begin{equation}
\textstyle
4\pi c \PiV^{\mbox{\tiny{field}}}
= \label{eq:PiMBLTP}
\DV\crprd\BV + \EV\crprd\HV - \EV\crprd\BV 
- \varkappa^{-2} \big(\nabla\cdot\EV\big)\Big(\nabla\crprd\BV - \frac{1}{c}\pddt\EV\Big).
\end{equation}

 As announced in \cite{KiePRD} and shown in \cite{KTZonBLTP}, BLTP electrodynamics is well-posed as a joint
initial value problem for the fields and the point charge, requiring initial data 
$\BV(0,\sV)$, $\DV(0,\sV)$, $\EV(0,\sV)$, $(\pddt\EV)(0,\sV)$ for the fields, and $\qV(0)$ and $\vV(0)$ for
the particle. 
 The data for $\BV$ and $\DV$ are constrained by the divergence equations, and  $\vV(0)$ by $|\vV(0)|<c$.

 For a charged particle released from rest in 
the constant applied electric field $\EV^{\mbox{\tiny{hom}}}$,
the initial data are
\begin{equation}
\qV(0)=\NullV
\qquad \mbox{and} \qquad
\vV(0)=\NullV.
\end{equation}
 The initial fields are the sum of the external constant electric field and the electrostatic field of the point charge,
thus we have
\begin{equation}\label{Dinit}
\DV(0,\sV)\equiv \EV^{\mbox{\tiny{hom}}} +\ee \frac{\sV}{|\sV|^3}
\end{equation}
and
\begin{equation}\label{Einit}
\EV(0,\sV)
\equiv \EV^{\mbox{\tiny{hom}}} + \ee \frac{1 - (1+\varkappa |\sV|)e^{-\varkappa |\sV|}}{|\sV|^2}\frac{\sV}{|\sV|},
\end{equation}
{in addition to the initial conditions}
 $\big(\partial_t \EV\big)(0,\sV)\equiv \NullV$, as well as $\BV(0,\sV) \equiv \NullV$.

%%%%%%%%%%%%%%%%%%%%%%%%%%%%%%%%%%%%%%%%%%%%%%%%%%%%%%%%%%%%%%
%%%%%%%%%%%%%%%%%%%%%%%%%%%%%%%%%%%%%%%%%%%%%%%%%%%%%%%%%%%%%%
%%%%%%%%%%%%%%%%%%%%%%%%%%%%%%%%%%%%%%%%%%%%%%%%%%%%%%%%%%%%%%%%%%%%
%%%%%%%%%%%%%%%%%%%%%%%%%%%%%%%%%%%%%%%%%%%%%%%%%%%%%%%%%%%%%%
     \section{Reduction to a Volterra integral equation for the particle acceleration} %\vspace{-10pt}
%%%%%%%%%%%%%%%%%%%%%%%%%%%%%%%%%%%%%%%%%%%%%%%%%%%%%%%%%%%%%%%%%%%%
%%%%%%%%%%%%%%%%%%%%%%%%%%%%%%%%%%%%%%%%%%%%%%%%%%%%%%%%%%%%%%
%%%%%%%%%%%%%%%%%%%%%%%%%%%%%%%%%%%%%%%%%%%%%%%%%%%%%%%%%%%%%%
%%%%%%%%%%%%%%%%%%%%%%%%%%%%%%%%%%%%%%%%%%%%%%%%%%%%%%%%%%%%%% 

 For the initial data of our problem the electromagnetic fields outside the forward light cone of 
the initial location of the particle remain precisely the electrostatic fields, i.e., the magnetic field $\HV$ 
and the magnetic induction field $\BV$ vanish, while the electric displacement field $\DV(t,\sV)$ is given by \refeq{Dinit}
and the electric field $\EV(t,\sV)$ is given by \refeq{Einit}, for all $t\geq 0$.

 Inside the forward light cone of the initial particle location, 
{but excluding the world line of the particle itself,}
the fields $\DV(t,\sV)$ and $\HV(t,\sV)$ are for all $t\geq 0$ given by
$\DV = \EV^{\mbox{\tiny{hom}}} + \DLWr$ \&\ $\HV =\HLWr$, with 
(the acceleration vector of the point charge is highlighted in {\color{red}red})
\begin{alignat}{1}
\hskip-.6truecm
{\DLWr(t,\sV)} &=\label{eq:LWsolD}
  \ee \frac{c^2-|\vV|^2}{|\sV-\qV|^2} 
\frac{{c\nV(\qV,\sV)}_{\phantom{!\!}}-{\vV}}{\bigl(\textstyle{c-\nV(\qV,\sV)\cdot {\vV}}\bigr)^{\!3}} 
\Biggl.\Biggr|^{\mathrm{ret}}
+\ee 
\frac{\nV(\qV,\sV)\crprd
\bigl[\bigl(c\nV(\qV,\sV)_{\phantom{!\!}}-{\vV}\bigr)\crprd{\color{red}\aV}\bigr]}{{|\sV-\qV|}
\bigl(\textstyle{c-\nV(\qV,\sV)\cdot {\vV}}\bigr)^{\!3}}
\Biggl.\Biggr|^{\mathrm{ret}},\hskip-1truecm
\\
\hskip-1truecm
{\HLWr(t,\sV)} 
&= \label{eq:LWsolH}
        \nV(\qV,\sV)|^{_{\mathrm{ret}}}\crprd {\DLWr(t,\sV)}
\, \vspace{-10pt}
\end{alignat}
the retarded Li\'enard--Wiechert fields \cite{JacksonBOOKb}. 
 Here, $\nV(\qV,\sV) = \frac{\sV-\qV}{|\sV-\qV|}$ is a \emph{normalized} vector from $\qV$ to $\sV$, and
{the notation ``$|^{\mathrm{ret}}$'' means that the quantities $\qV$ and $\vV$ and ${\color{red}\aV}$ 
to the left of ``$|^{\mathrm{ret}}$'' are to be evaluated at the retarded time $t^{\mathrm{ret}}$, 
with $t^{\mathrm{ret}}(t,\sV)$ being defined implicitly by $c(t-t^{\mathrm{ret}}) = |\sV-\qV(t^{\mathrm{ret}})|$;
thus, for instance, 
$(\qV,\vV,{\color{red}\aV})|^{\mathrm{ret}}= (\qV(t^{\mathrm{ret}}),\vV(t^{\mathrm{ret}}),{\color{red}\aV}(t^{\mathrm{ret}}))$.
 Inside the initial forward light cone, $0< t^{\mathrm{ret}} <t$.
The terms $\propto{\color{red}\aV}$ in (\ref{eq:LWsolD}) and (\ref{eq:LWsolH}) account for radiation emitted into the 
far-field $(|\sV|\gg |\qV|$).}

 Note that the electromagnetic Li\'enard--Wiechert fields $\HLWr$ and $\DLWr$ exhibit both
a $\propto 1/r^2$ singularity and a $\propto 1/r$ singularity, where $r$ denotes $|\sV-\qV(t)|$; they each
have a directional singularity at the location of the  point charge source, too.

{Analogously the fields $\BV$ and $\EV$ for a moving point charge can be expressed \cite{KiePRD}; 
cf. \cite{Lazar}.
 Inside and on} the forward light cone of the initial particle location, 
{but excluding the world line of the particle itself,}
the MBLTP field solutions $\BV(t,\sV)$ and $\EV(t,\sV)$ for $t\geq 0$ are given by 
$\BV=\BV_0+ \BV_1$ and $\EV = \EV_0 + \EV_1$, 
with $\BV_0\equiv\NullV$ and $\EV_0\equiv \EV^{\mbox{\tiny{hom}}}$, and
(see \cite{KiePRD})
\begin{alignat}{1}
\hspace{-20pt}
{\EV_1(t,\sV)} =\; \nonumber
 & \ee \varkappa^2\Bigl(\frac{1 - (1+\varkappa |\sV|)e^{-\varkappa |\sV|}}{\varkappa^2|\sV|^2} -\tfrac12 \Bigr)\frac{\sV}{|\sV|} + 
\ee \varkappa^2 \int_{0}^{ct-|\sV|}
\tfrac{J_2\!\bigl(\varkappa\sqrt{c^2(t-t')^2-|\sV|^2}\bigr)}{{c^2(t-t')^2-|\sV|^2}^{\phantom{n}} }
 \sV \drm{(ct')} +
\\  \label{eq:EjsolMBLTP} 
 & \ee \varkappa^2\tfrac12\tfrac{\nV(\qV^{},\sV)-{\vV^{}}/{c}}{1-\nV(\qV^{},\sV)\cdot{\vV}/{c}}
\Big|^{\mathrm{ret}} \, - \\ \nonumber
 &
\ee \varkappa^2 \int_{0}^{t^\mathrm{ret}(t,\sV)}
\tfrac{J_2\!\bigl(\varkappa\sqrt{c^2(t-t')^2-|\sV-\qV(t')|^2}\bigr)}{{c^2(t-t')^2-|\sV-\qV(t')|^2}^{\phantom{n}} }
\left(\sV-\qV^{}(t') - \vV^{}(t')(t-t')\right)c\drm{t'} , \\ 
\hspace{-20pt}
{\BV_1(t,\sV)} =\; \label{eq:BjsolMBLTP}
&\ee \varkappa^2 \tfrac{1}{2} 
\tfrac{{\color{black}\vV^{}\crprd\nV(\qV^{},\sV)/c}}{1-\nV(\qV^{},\sV)\cdot{\vV}/{c}}
\Big|^{\mathrm{ret}}
\, -  \\ \nonumber
 &
 \ee \varkappa^2 \int_{0}^{t^\mathrm{ret}(t,\sV)}
\tfrac{J_2\!\bigl(\varkappa\sqrt{c^2(t-t')^2-|\sV-\qV(t')|^2 }\bigr)}{{c^2(t-t')^2-|\sV-\qV(t')|^2}^{\phantom{n}} }
{\vV^{}(t')}\crprd \left(\sV-\qV^{}(t') 
\right)\drm{t'} .
\end{alignat}
 The fields $\BV(t,\sV)$ and $\EV(t,\sV)$ 
are globally bounded in $\sV$ for each $t$, and away from the point charge they are Lipschitz-continuous in $\sV$, including
across the initial forward light cone. 

 Similarly (see \cite{CarKie}),
\begin{alignat}{1}
\qquad\qquad\nabla\cdot\EV(t,\sV) =\; \nonumber
 & \ee \varkappa^2 \frac{e^{-\varkappa |\sV|} - 1}{|\sV|}
+ 
\ee \varkappa^3 \int_{0}^{ct-|\sV|}
\tfrac{J_1\!\bigl(\varkappa\sqrt{c^2(t-t')^2-|\sV|^2}\bigr)}{\sqrt{c^2(t-t')^2-|\sV|^2}^{\phantom{n}} }
 \drm{(ct')} +
\\  \label{eq:LWsolPHI}
& \ee^{}\varkappa^2 
\tfrac{1}{\bigl(1-\nV(\qV,\sV)\cdot {\vV^{}}/{c}\bigr)}
\tfrac{1}{|\sV-\qV|}
\Bigl.\Bigr|^{\mathrm{ret}}
- \\ \nonumber
&
\ee^{} \varkappa^3 \int_{0}^{t^\mathrm{ret}(t,\sV)}
\tfrac{J_1\!\bigl(\varkappa\sqrt{c^2(t-t')^2-|\sV-\qV(t')|^2}\bigr)}{\sqrt{c^2(t-t')^2-|\sV-\qV(t')|^2}^{\phantom{n}} }
c \drm{t'} ,
\end{alignat}
and
\begin{alignat}{1}
\big(\nabla\crprd\BV - {\textstyle{\frac{1}{c}\pddt}}\EV\big)(t,\sV) 
=\; \label{eq:LWsolA}
& \ee^{}\varkappa^2\tfrac{1}{\bigl(1-\nV(\qV,\sV)\cdot {\vV^{}}/{c}\bigr)}
\tfrac{1}{|\sV-\qV|}\tfrac{\vV^{}}{c}
\Bigl.\Bigr|^{\mathrm{ret}}
- \\ \nonumber
 & \ee^{} \varkappa^3  \int_{0}^{t^\mathrm{ret}(t,\sV)} 
\tfrac{J_1\!\bigl(\varkappa\sqrt{c^2(t-t')^2-|\sV-\qV(t')|^2}\bigr)}{\sqrt{c^2(t-t')^2-|\sV-\qV(t')|^2}^{\phantom{n}} }
\vV^{}(t')\drm{t'} .
\end{alignat}

 With the help of the solution formulas for the fields, given the motions, 
the electromagnetic force of the MBLTP field on its point charge source can be expressed as a functional over
the maps $t\mapsto (\qV(t),\vV(t),{\color{red}\aV}(t))$.
 Namely, since each electromagnetic field component is the sum of a vacuum field and a sourced field, the bilinear 
$\PiV^{\mbox{\tiny{field}}}$ decomposes into a sum of three types of terms: the vacuum-vacuum terms, the source-source 
terms, and the mixed vacuum-source terms.
 In our problem the vacuum field is $\EV^{\mbox{\tiny{hom}}}$; it does not contribute to $\PiV^{\mbox{\tiny{field}}}$,
but appears separately at rhs\refeq{eq:POINCAREforce}.
 As explained in \cite{KiePRD}, this term is not put in by hand but is a contribution to the momentum 
balance due to a surface integral at ``$|\sV|=\infty$.''
 Hence the only contribution to rhs\refeq{eq:POINCAREforce} from $\PiV^{\mbox{\tiny{field}}}$ is the 
source-source contribution, a self-field force in BLTP electrodynamics. 
 Thus, \refeq{eq:POINCAREforce} is given by 
\begin{equation}\label{eq:totalF}
\fV(t) 
=
 \ee \EV^{\mbox{\tiny{hom}}}
+
\fV^{\mbox{\tiny{self}}}[\qV,\vV;{\color{red}\aV}](t),
\end{equation} 
where $\ee \EV^{\mbox{\tiny{hom}}}$ is the Lorentz force evaluated with the vacuum field (i.e., a ``test particle contribution'' to the total force), and (after taking advantage of hyperbolicity; cf. \cite{KiePRD})
\begin{alignat}{2}\label{eq:selfFa}
\hspace{-20pt}
\fV^{\mbox{\tiny{self}}}[\qV,\vV;{\color{red}\aV}](t)
 \equiv &  - \frac{\drm}{\drm{t}} \displaystyle\int_{B_{ct}(\qV_0)}
\Bigl( \PiV^{\mbox{\tiny{field}}}_{\mbox{\tiny{source}}}(t,\sV) - 
\PiV^{\mbox{\tiny{field}}}_{\mbox{\tiny{source}}}(0,\sV-\qV_0-\vV_{\!0}t)\Bigr) d^3{s}  \\
 =& - \frac{\drm}{\drm{t}} \displaystyle\int_{B_{ct}(\qV_0)}
 \PiV^{\mbox{\tiny{field}}}_{\mbox{\tiny{source}}}(t,\sV)  d^3{s}  ,\label{selfFb}
\end{alignat}
with $\PiV^{\mbox{\tiny{field}}}_{\mbox{\tiny{source}}}$ given by \refeq{eq:PiMBLTP} 
with $(\BV,\DV-\EV^{\mbox{\tiny{hom}}},\EV-\EV^{\mbox{\tiny{hom}}},\HV)$ in place of $(\BV,\DV,\EV,\HV)$, and where ${B_{ct}(\qV_0)}$ is the ball of radius $ct$ centered at $\qV_0$. 
 To go from \refeq{eq:selfFa} to 
 \refeq{selfFb} we made use of the initial data $\qV_0=\NullV$ and $\vV_0=\NullV$, and 
$\PiV^{\mbox{\tiny{field}}}_{\mbox{\tiny{source}}}(0,\sV)\equiv \NullV$.

 The following merits emphasis: The r.h.s.(\ref{selfFb}) contains the acceleration ${\color{red}\aV}$ only linearly 
through the contributions $\propto \DV$ and $\propto \HV$ in  $\PiV^{\mbox{\tiny{field}}}_{\mbox{\tiny{source}}}$ 
given by \refeq{eq:PiMBLTP} 
with $(\BV,\DV-\EV^{\mbox{\tiny{hom}}},\EV-\EV^{\mbox{\tiny{hom}}},\HV)$ in place of $(\BV,\DV,\EV,\HV)$.
 Therefore the equation of motion for the point charge can be viewed as 
a linear Volterra integral equation for $t\mapsto{\color{red}\aV}(t)$, given $t\mapsto\qV(t)$ and $t\mapsto\vV(t)$ (even without
demanding that $\vV(t) =\dot\qV(t)$), viz.
\begin{equation}\label{Volterra}
\forall t:\ 
{\color{red}\aV}(t)= W[\vV]\cdot\left(\eEL \EV^{\mbox{\tiny{hom}}}+\fV^{\mbox{\tiny{self}}}[\qV,\vV;{\color{red}\aV}]\right)(t).
\end{equation}
 Here,
\begin{equation}
 W[\vV]
:= \label{eq:FtoAmapINv}
\textstyle
\frac{1}{\mbare}\sqrt{1-\frac{|\vV|^2}{c^2}}
\left[\ID - \frac{1}{c^2}\vV\otimes \vV\right],
\end{equation}
which for motion along $\EV^{\mbox{\tiny{hom}}}$ simplifies to
\begin{equation}
 W[\vV]
:= \label{eq:FtoAmapINvAGAIN}
\textstyle
\frac{1}{\mbare}\left(1-\frac{|\vV|^2}{c^2}\right)^{3/2}\ID .
\end{equation}
 In \cite{KTZonBLTP} we show that the Volterra equation can be uniquely solved to yield ${\color{red}\aV}$ as a nonlinear
functional of (a priori unrelated) curves $t\mapsto\qV(t)$ and $t\mapsto\vV(t)$. 
 Finally setting $\vV(t) = \dot\qV(t)$ and ${\color{red}\aV}(t) = \ddot\qV(t)$, 
 the solution to the Volterra integral equation poses a second-order differential equation 
initial value problem for $\qV(t)$ without the infamous $\dddot\qV(t)$ term of the Abraham--Lorentz--Dirac equation.

 To compute $\vV(t)$ and $\qV(t)$, 
the integral at r.h.s.(\ref{selfFb}) will be rewritten with the help of retarded spherical coordinates (see our Appendix),
then expanded in powers of $\varkappa$, and then all terms up to $O(\varkappa^4)$ included will be evaluated in terms of 
$\qV(t)$, $\vV(t)$, and in terms of integrals over time involving $t\mapsto\qV(t)$, $t\mapsto\vV(t)$, and 
$t\mapsto{\color{red}\aV}(t)$. 
 While these evaluations are daunting, the eventual outcomes up to order $\varkappa^3$ included are surprisingly simple;
see \cite{CarKie}. 
 Our new contribution, the $O(\varkappa^4)$ term, {is only slightly more complicated and 
readily amenable to numerical treatment on a computer.}
\vspace{-0.5truecm}

%%%%%%%%%%%%%%%%%%%%%%%%%%%%%%%%%%%%%%%%%%%%%%%%%%%%%%%%%%%%%%
%%%%%%%%%%%%%%%%%%%%%%%%%%%%%%%%%%%%%%%%%%%%%%%%%%%%%%%%%%%%%%
%%%%%%%%%%%%%%%%%%%%%%%%%%%%%%%%%%%%%%%%%%%%%%%%%%%%%%%%%%%%%%%%%%%%
                \section{Small-$\varkappa$ expansion of the Volterra equation}\label{sec:MBLTPkappaSMALL}\vspace{-0.2truecm}
%%%%%%%%%%%%%%%%%%%%%%%%%%%%%%%%%%%%%%%%%%%%%%%%%%%%%%%%%%%%%%%%%%%%
%%%%%%%%%%%%%%%%%%%%%%%%%%%%%%%%%%%%%%%%%%%%%%%%%%%%%%%%%%%%%%
%%%%%%%%%%%%%%%%%%%%%%%%%%%%%%%%%%%%%%%%%%%%%%%%%%%%%%%%%%%%%%

 The formal power series expansion about $\varkappa=0$ of  $\fV^{\mbox{\tiny{self}}}$ is
 $\fV^{\mbox{\tiny{self}}}[\qV,\vV;{\color{red}\aV}](t) = \sum\limits_{n=0}^\infty \FV^{(n)}_0[\qV,\vV;{\color{red}\aV}](t)$,
where $\FV^{(n)}_0 \propto \varkappa^n$;
 the subscript ${}_0$ at $\FV^{(n)}_0$ indicates that we are expanding about $\varkappa=0$. 
 It is manifest that the terms $O(\varkappa^0)$ and $O(\varkappa^1)$ vanish identically, so we need to discuss terms
$O(\varkappa^n)$ for $n\geq 2$.
 Several of the spherical integrations can been carried out explicitly in terms of well-known functions.
 In particular, the contributions $\FV^{(2)}_0$, $\FV^{(3)}_0$, and $\FV^{(4)}_0$ can be computed explicitly
as expressions in $\qV$ and $\vV$ at time $t$, and time integrals over algebraic expressions of $\qV$ and $\vV$ and
${\color{red}\aV}$. 
 Remarkably, the integrations involving ${\color{red}\aV}$ can be carried out and reduced to 
time integrals over algebraic expressions of only $\qV$ and $\vV$. 

 In the ensuing subsections \ref{sec:MBLTPkappaSQR}, \ref{sec:MBLTPkappaTHREE}, and \ref{sec:MBLTPkappaFOUR}
 we will temporarily suppress the argument $[\qV,\vV;{\color{red}\aV}]$ from the $\FV^{(k)}$.\vspace{-0.5truecm}
%%%%%%%%%%%%%%%%%%%%%%%%%%%%%%%%%%%%%%%%%%%%%%%%%%%%%%%%%%%%%%
%%%%%%%%%%%%%%%%%%%%%%%%%%%%%%%%%%%%%%%%%%%%%%%%%%%%%%%%%%%%%%%%%%%%
                \subsection{The self-force at $O(\varkappa^2)$}\label{sec:MBLTPkappaSQR} \vspace{-7pt}
%%%%%%%%%%%%%%%%%%%%%%%%%%%%%%%%%%%%%%%%%%%%%%%%%%%%%%%%%%%%%%%%%%%%
%%%%%%%%%%%%%%%%%%%%%%%%%%%%%%%%%%%%%%%%%%%%%%%%%%%%%%%%%%%%%%
 
 The $O(\varkappa^2)$ contribution to the self-force has been computed in \cite{CarKie}
(see also the erratum in \cite{KiePRD}), and it turns out to be
\begin{alignat}{1}
 \FV^{(2)}_0(t) = \label{eq:SELFforceATorderTWOmerged}
 \, \NullV.
\end{alignat}
 In this problem of straight line motion in a constant external electric field, with the particle starting from rest,
the BLTP radiation-reaction force \emph{vanishes identically} at $O(\varkappa^2)$.
\vspace{-0.5truecm}

%%%%%%%%%%%%%%%%%%%%%%%%%%%%%%%%%%%%%%%%%%%%%%%%%%%%%%%%%%%%%%
%%%%%%%%%%%%%%%%%%%%%%%%%%%%%%%%%%%%%%%%%%%%%%%%%%%%%%%%%%%%%%%%%%%%
                \subsection{The self-force at $O(\varkappa^3)$}\label{sec:MBLTPkappaTHREE} \vspace{-7pt}
%%%%%%%%%%%%%%%%%%%%%%%%%%%%%%%%%%%%%%%%%%%%%%%%%%%%%%%%%%%%%%%%%%%%
%%%%%%%%%%%%%%%%%%%%%%%%%%%%%%%%%%%%%%%%%%%%%%%%%%%%%%%%%%%%%%

 The $O(\varkappa^3)$ contribution to the radiation-reaction force for small $\varkappa$ has been computed in \cite{CarKie}.
 It reads
\begin{alignat}{1}
\label{eq:F3}
 \FV^{(3)}_0(t) = 
  - \frac13   \varkappa^3 \eEL^2\qV(t) .
\end{alignat}

 This is a very surprising result: the $O(\varkappa^3)$ term of the radiation-reaction force in our initial value problem 
is a harmonic oscillator force!
 This result relies on the particular setup of the initial data and the geometry of the problem, but not more. 
\vspace{-0.5truecm}
%%%%%%%%%%%%%%%%%%%%%%%%%%%%%%%%%%%%%%%%%%%%%%%%%%%%%%%%%%%%%%
%%%%%%%%%%%%%%%%%%%%%%%%%%%%%%%%%%%%%%%%%%%%%%%%%%%%%%%%%%%%%%%%%%%%
                \subsection{The self-force at $O(\varkappa^4)$}\label{sec:MBLTPkappaFOUR}\vspace{-7pt}
%%%%%%%%%%%%%%%%%%%%%%%%%%%%%%%%%%%%%%%%%%%%%%%%%%%%%%%%%%%%%%%%%%%%
%%%%%%%%%%%%%%%%%%%%%%%%%%%%%%%%%%%%%%%%%%%%%%%%%%%%%%%%%%%%%%

 The $O(\varkappa^4)$ contribution to the radiation-reaction force for small $\varkappa$ is computed in this paper.
 It reads
\begin{equation}
\label{eq:F4}
\hspace{-10pt}
\FV^{(4)}_0(t)=
\frac{1}{4}\varkappa^4 e^{2} \!\int_{0}^{t}\qV(t^{r} )c\drm{t}^{r} .
\end{equation}
As mentioned in the introduction, the technical details for obtaining this result are supplied in the appendix.

 Also the $O(\varkappa^4)$ term is surprisingly simple, though not quite as simple as the $O(\varkappa^3)$ term.

\vspace{-0.5truecm}

%%%%%%%%%%%%%%%%%%%%%%%%%%%%%%%%%%%%%%%%%%%%%%%%%%%%%%%%%%%%%%
%%%%%%%%%%%%%%%%%%%%%%%%%%%%%%%%%%%%%%%%%%%%%%%%%%%%%%%%%%%%%%%%%%%%
%%%%%%%%%%%%%%%%%%%%%%%%%%%%%%%%%%%%%%%%%%%%%%%%%%%%%%%%%%%%%%
\section{Integro-differential equation for the particle position to $O(\varkappa^4)$ included, and numerical solutions}
%%%%%%%%%%%%%%%%%%%%%%%%%%%%%%%%%%%%%%%%%%%%%%%%%%%%%%%%%%%%%%%%%%%%
%%%%%%%%%%%%%%%%%%%%%%%%%%%%%%%%%%%%%%%%%%%%%%%%%%%%%%%%%%%%%%
%%%%%%%%%%%%%%%%%%%%%%%%%%%%%%%%%%%%%%%%%%%%%%%%%%%%%%%%%%%%%%
\vspace{-0.2truecm}

 When all ${\color{red}\aV}$ terms up to $O(\varkappa^4)$ included are
converted through time-integration into expressions that involve only $\qV$
and $\vV$, the Volterra equation for ${\color{red}\aV}$ becomes an integro-differential equation for $\qV$ of
second order.
 Writing now $\FV^{(k)}_0[\qV,\vV](t)$ instead of $\FV^{(k)}_0[\qV,\vV;{\color{red}\aV}]$ 
to highlight this elimination of ${\color{red}\aV}$ in favor of $\qV$ and $\vV$
in the self-force terms, the integro-differential equation for $\qV(t)$ reads 
\begin{equation}\label{VolterraATorderFOUR}
\ddot\qV(t)= 
\textstyle
\frac{1}{\mbare}\bigl(1-\frac{1}{c^2}|\vV(t)|^2\bigr)^{\frac32}\left(\eEL \EV^{\mbox{\tiny{hom}}} 
+ \FV^{(3)}_0[\qV,\vV](t)
+ \FV^{(4)}_0[\qV,\vV](t)
\right),
\end{equation}
with $\FV^{(3)}_0[\qV,\vV]$ and $\FV^{(4)}_0[\qV,\vV](t)$ given by r.h.s.(\ref{eq:F3}) and r.h.s.(\ref{eq:F4}), 
respectively.

 Incidentally, if we drop the $O(\varkappa^4)$ term we obtain the $O(\varkappa^3)$ approximation discussed in 
\cite{CarKie}.
 For positive $\mbare$ it is equivalent to the problem of special-relativistic test particle motion in a harmonic 
oscillator potential, but featuring anharmonic time-periodic solutions due to the relativistic mass change.
  For negative $\mbare$ it is equivalent to the problem of special-relativistic positive mass test particle motion 
in a ``harmonic repeller'' potential, featuring run-away solutions whose speed monotonically approaches the speed 
of light.
 In either case, the motions conserve the effective total particle energy 
\begin{equation}\label{EFFECTIVEenergy}
U = \frac{\mbare c^2}{\sqrt{ 1- \frac{1}{c^2}|\vV|^2}} 
- \eEL  \EV^{\mbox{\tiny{hom}}}\cdot\qV  +\tfrac16  \eEL^2 \varkappa^3|\qV|^2.
\end{equation}
 Since $\vV(t) = \frac{d\qV(t)}{dt}$, and since we have straight line motion, i.e. $\vV\|\qV$ for all $t$ for 
which both vectors are non-vanishing, this is an implicit first-order differential equation for $t(\qV)$ 
(in the usual simplified notation). 
 Eq. \ref{EFFECTIVEenergy} can easily be converted into an explicit first-order differential equation for $t(\qV)$ that
then can be solved by quadrature.

 The study \cite{CarKie} left open whether the mathematical validity of the $O(\varkappa^3)$ approximation to the
untruncated Volterra equation is restricted to short times $\varkappa ct\ll 1$ or whether such
periodic motion over longer times for $\mbare>0$, respectively monotonic run-away solutions for $\mbare<0$, 
are genuine features of BLTP electrodynamics as long as $\varkappa e^2/|\mbare| c^2\ll 1$.
 If the $O(\varkappa^3)$ approximation is faithful provided $\frac{\varkappa e^2 }{ |\mbare| c^2} \ll 1$, then BLTP 
electrodynamics would have to be purged from the list of contenders for a realistic classical electrodynamics.

 By now a verdict is in. 
Fig.s~1--4 below show solutions of (\ref{VolterraATorderFOUR}) computed numerically after converting (\ref{VolterraATorderFOUR}) into a system of first-order differential equations by defining the unit vector $\be$ as $\EV^{\mbox{\tiny{hom}}}=: |\EV^{\mbox{\tiny{hom}}}|\be$, and then defining the variables $q(t):=\be\cdot\qV(t)$, $v(t):=\be\cdot\vV(t)$, $X(t) := \be\cdot \int_{0}^{t}\qV(t^{r} )c dt^{r}$, using that $\dot{q}(t) = v(t)$.
 This equivalent first order ODE system thus reads
\begin{align}
\dot{q}(t) & = v(t),\\
\dot{v}(t) & = \frac{1}{\mbare}\left(1-\beta(t)^2\right)^{\frac32}\left(\eEL |\EV^{\mbox{\tiny{hom}}}|
 -\frac13  \varkappa^3 \eEL^2 q(t) + \frac{1}{4}\varkappa^4 e^{2} X(t)\right)  ,
\\
\dot{X}(t) & = cq(t),
\end{align}
with $\beta(t) = \frac1c \vV(t)\cdot\be$.
 We supplemented this ODE system with vanishing initial data for $X(t)$, and also for $q(t)$ and $v(t)$.

 The solutions to the $O(\varkappa^4)$ approximation that we computed numerically confirm that the ``small $\varkappa$'' 
expansion is mathematically applicable for ``small times $t$'', viz. when  $\varkappa ct\ll 1$, 
with the motions at $O(\varkappa^0)$, $O(\varkappa^3)$, and $O(\varkappa^4)$ barely distinguishable.
 However, truncation after $O(\varkappa^3)$ does not yield mathematically accurate approximations to 
the true BLTP motions over arbitrarily long times even if $\frac{\varkappa e^2}{|\mbare| c^2}\ll~1$. 
 Should this expansion converge to the actual BLTP motions also over longer times
provided $\varkappa e^2/|\mbare| c^2 \ll 1$, then reasonable accuracy will require to go to much higher powers of
$\varkappa$.  
 In the following we distinguish $\mbare>0$ and $\mbare<0$, beginning with $\mbare>0$.

 For the first figure the parameter values are the same as in Fig.~1 of \cite{CarKie}. %
 In Figs.~2 and 3 we keep $\mbare>0$ but vary the parameter values compared to \cite{CarKie}. 
 Figures 1--3 all reveal that the $O(\varkappa^4)$ motion is faster than the $O(\varkappa^3)$ motion for late enough
times, and eventually even faster than the test particle motion, {yet}
the resolution is not fine enough to reveal what happens for very short time spans after release of the particle.
 This is readily worked out by expanding the self-force about $t=0$, noting that the leading order contribution is 
the test particle motion $\qV^{\mbox{\tiny{test}}}(t) = \frac12 \frac{e}{\mbare}\EV^{\mbox{\tiny{hom}}}t^2$,
which yields
\begin{alignat}{1}
\label{selfFexpandedNEARzeroA}
\Big( \FV^{(3)}_0 +  \FV^{(4)}_0\Big)[\qV,\vV](t) & = 
  - \frac13  \eEL^2 \varkappa^3\qV(t) 
  + \frac14  \eEL^2 \varkappa^4  \!\int_{0}^{t} \qV(t^r) c \drm{t}^{r}  \\
\label{selfFexpandedNEARzeroB}
& = \eEL\EV^{\mbox{\tiny{hom}}}\frac{\varkappa \eEL^2 }{\mbare c^2} \Big(- \frac16  (\varkappa c t)^2
+
 \frac{1}{24} (\varkappa ct)^3
+ O\big((\varkappa c t)^4\big)\Big).
\end{alignat}
 This demonstrates that for very short time spans after release of the particle, 
the $O(\varkappa^4)$ motion is faster than the $O(\varkappa^3)$ motion, 
but both are slower than the radiation-free test particle motion. 
 {Note that in (\ref{selfFexpandedNEARzeroB}) the two ``small $\varkappa$'' parameters, 
viz. $\frac{\varkappa \eEL^2 }{\mbare c^2}$ and $\varkappa ct$, are exhibited separately to highlight
their respective significances.} 

\begin{figure}[H]
  \includegraphics[width = 8truecm,scale=3]{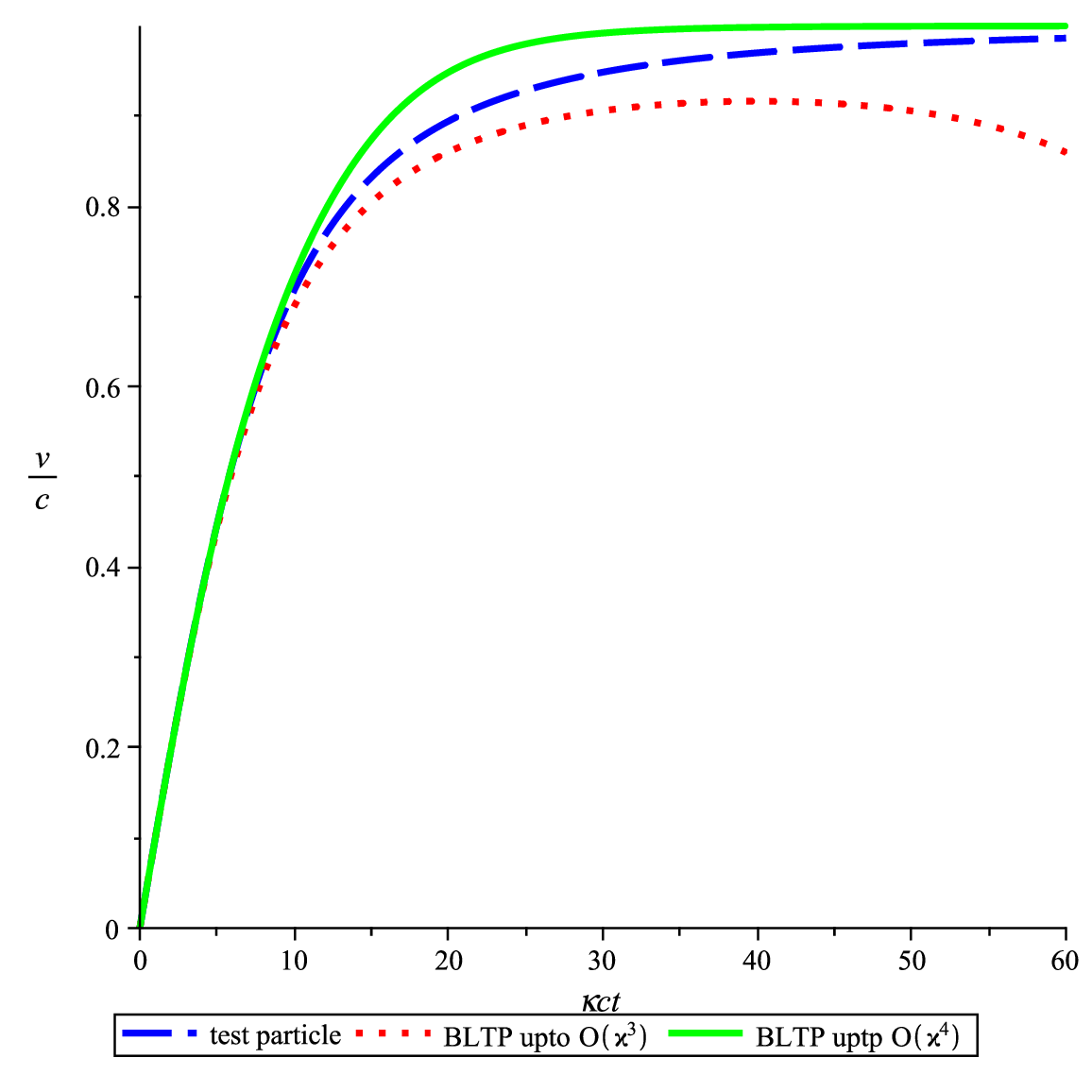} \vspace{-.3truecm}
\caption{
\footnotesize{The velocity of a point charge, starting from rest in a constant applied electrostatic 
 field $\EV^{\mbox{\tiny{hom}}}=10e\varkappa^2$, vs. time, as per test particle theory (dashed curve), resp. BLTP electrodynamics
 with radiation-reaction included to $O(\varkappa^3)$ (dotted curve), resp. to
 $O(\varkappa^4)$ (continuous curve), when $\varkappa e^2/\mbare c^2 =0.01$.
 The period of the velocity of the $O(\varkappa^3)$ BLTP motion is $\varkappa c T = 160$. 
The test particle's velocity and the BLTP particle's $O(\varkappa^4)$ velocity asymptote to $c$.}
}
\end{figure}

\begin{figure}[H]
  \includegraphics[width = 8truecm,scale=3]{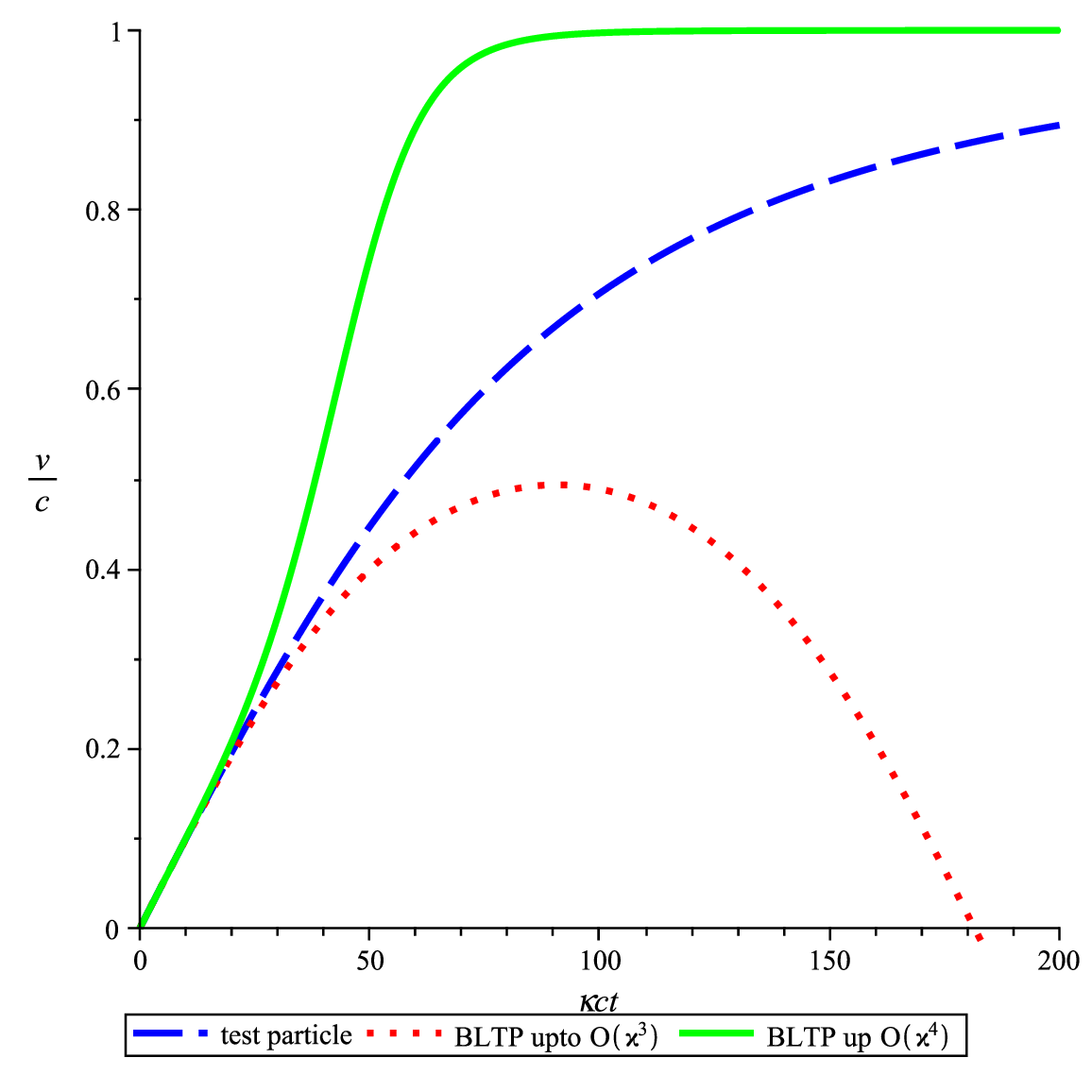} \vspace{-.3truecm}
\caption{
\footnotesize{Same as Fig.~1, but now for parameter values $\EV^{\mbox{\tiny{hom}}}=10e\varkappa^2$
and $\varkappa e^2/\mbare c^2 =0.001$.}
}
\end{figure}

\begin{figure}[H]
  \includegraphics[width = 8truecm,scale=3]{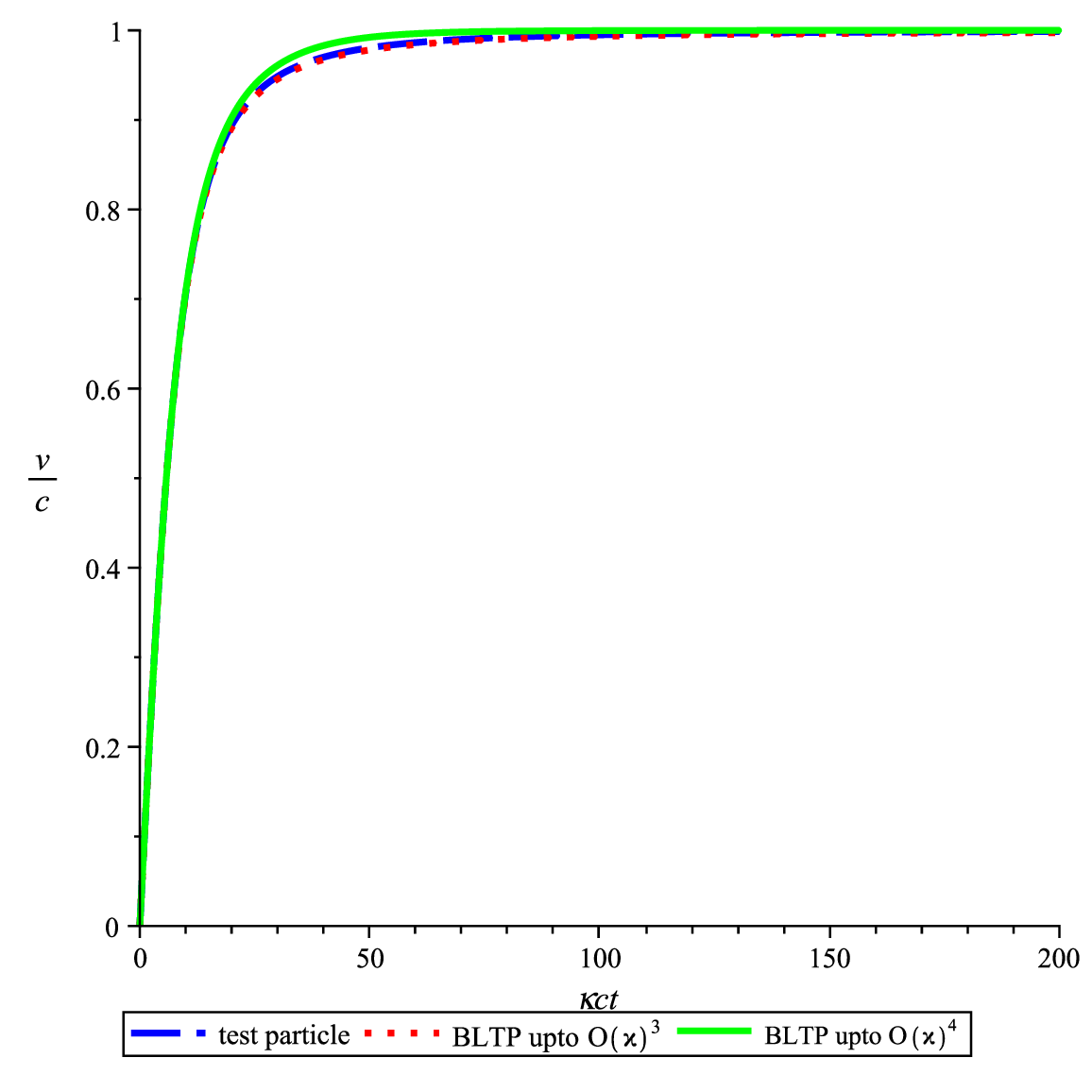} \vspace{-.3truecm}
\caption{
\footnotesize{Same as Fig.~1, but now for parameter values $\EV^{\mbox{\tiny{hom}}}=100e\varkappa^2$
and $\varkappa e^2/\mbare c^2 =0.001$.}
}
\end{figure}

 For short times $\varkappa ct \ll 1$ 
Fig.s~1--3 indicate rapid convergence of the power series expansion to actual BLTP
motions, and for such short times the computed motions are physically reasonable, with radiation-reaction effects acting against
the accelerating force of the applied electric field $\EV^{\mbox{\tiny{hom}}}$.
 Interestingly, one should note that the test particle motion and the motions 
that include all self-force effects to $O(\varkappa^3)$, respectively
to $O(\varkappa^4)$, are barely distinguishable even until some times outside the $\varkappa ct \ll 1$ regime:
in Fig.~1 good agreement lasts up to $\varkappa ct <5$ (approximately),
while for the parameters in Fig.~2 and Fig.~3 this appears to be the case as long as $\varkappa ct < 10$ (roughly). 
 This hints at a convergence of the expansion in powers of $\varkappa$ when $\frac{\varkappa e^2}{\mbare c^2}\ll 1$ also
for much larger times than those satisfying $\varkappa ct \ll 1$.

 However, even if convergence holds for all times when $\frac{\varkappa e^2}{\mbare c^2}\ll 1$, 
based on our results one cannot decide whether the $O(\varkappa^4)$ motions closely approximate the actual BLTP motions.
 Therefore, to get to interestingly large times one would need to expand to much higher order. 
 What is clear from Fig.s~1 and 2 is that even though the parameter $\frac{\varkappa e^2}{\mbare c^2}\ll 1$, the
motions computed to $O(\varkappa^3)$ included do not closely approximate the motions computed to $O(\varkappa^4)$ included
when $\varkappa ct$ becomes too big, roughly when $\varkappa ct >5$ (for Fig.~1), respectively $\varkappa ct > 10$ (Fig.~2). 
 Fig.~3 is not as revealing, for the motions are temporarily not very close (though close), but do look very close again for 
$\varkappa ct > 100$. 
 Yet this continues to be true only
for the comparison between test particle and $O(\varkappa^4)$ motion; recall that the $O(\varkappa^3)$ motion is periodic, so
eventually one would see serious differences between $O(\varkappa^4)$ and $O(\varkappa^3)$ motions.

 So far we have addressed the mathematical significance of the ``small-$\varkappa$'' regime.
  The physical meaning of the ``small $\varkappa$'' regime for positive bare mass $\mbare$ is as follows.
 In scattering experiments, when the participating charged particles are moving
asymptotically freely and drag along their electrostatic fields (defined by a Lorentz transformation into the co-moving frame),
that electric field contributes to the total mass 
 $\mEL = \mbare + E^{\mbox{\tiny{field}}}/c^2$, with $E^{\mbox{\tiny{field}}} = \frac12\varkappa e^2$.
 The traditional interpretation is that $\mEL$ is to be identified with the mass of the empirical electron, which in
theory is a ``dressed electron'' with effective mass $\mEL>0$.
 The notion of dressing refers to the electric field that ``clothes'' a charge with bare mass and thus contributes
to the empirically accessible mass.
 
 Since $\frac12\varkappa e^2 >0$, the definition $\mEL = \mbare + E^{\mbox{\tiny{field}}}/c^2$ implies 
that $\mbare < \mEL$.
 Now empirically $\mEL>0$, and so, when $\mbare>0$, too, 
then the condition $\frac{\varkappa e^2}{\mbare c^2}\ll 1$ implies that $\mbare \approx \mEL$. 
 This in itself is perfectly reasonable, but it means that the electrostatic self-field energy only makes a 
small contribution to the total mass, and this can only happen if the effective ``size of the electron'' $\varkappa^{-1}$ 
in the BLTP model is huge --- in conflict with many empirical data other than the empirical mass. 

 Yet, when the Bopp length $\varkappa^{-1}$ is very small, as forced on us in particular
by demanding agreement of BLTP hydrogen data with empirical hydrogen spectral data \cite{CKP}, 
then the electrical self-energy $\frac12\varkappa e^2$ is huge, and then, to obtain the much smaller positive empirical mass, 
the bare mass needs to be negatively huge. 
 One then has $\frac12\varkappa e^2/\mbare c^2 \approx -1$.
 While this clearly violates the ``small $\varkappa$'' condition $\frac{\varkappa e^2}{|\mbare| c^2}\ll 1$,
for short time $\varkappa ct \ll 1$ we are still formally in a ``small-$\varkappa$'' regime in a different sense,
and so we expect rapid mathematical convergence of our power series expansion for such small times, hence insights into the 
radiation-reaction-corrected motions with negative bare mass.
 We have solved our system of $O(\varkappa^4)$ equations numerically for such a choice of parameters, see Fig.~4.
 And while for the positive bare mass regime we found by visual inspection that the motions seem to be accurately approximated
by the solutions to the truncations of the equations of motion even somewhat into the $\varkappa ct > 1$ region, 
presumably because we were also in the ``small $\varkappa$'' regime $\frac{\varkappa e^2}{\mbare c^2} \ll 1$, we
cannot reasonably expect the same to be true in the negative bare mass region 
when $\frac{\varkappa e^2}{|\mbare| c^2} \approx 2$. 
 Nevertheless we have solved the truncated equations of motion also for times $\varkappa ct >1$, 
just to see what happens. 
 The result is highly interesting and  shown in Fig.~4.

{Since in the BLTP initial value problem it is the bare mass that accounts for the particle's inertia
in the initial phase when a particle is released from rest in a static field, a negative-bare-mass particle is 
inevitably setting its sail in the opposite direction of the motion of a particle with positive bare mass that is
shown in Fig.~1--3.
 This is clearly seen in Fig.~4, where we also show the curve labeled \emph{charge soliton} that represents the usual textbook's
test particle motion 
of a dressed electron with positive effective mass $\mEL$, though we chose a convenient value for $\mEL$ for illustrative purposes. 
 For our terminlogy \emph{charge soliton} we refer to Spohn's book \cite{Spohn}.\footnote{Spohn and collaborators 
proved that these attract the long time 
adiabatic motion of the smeared out charged particles interacting with the Maxwell--Lorentz fields: the Abraham 
model of classical electron theory; respectively some quasi-relativistic version of the corresponding Lorentz
model.}
A charge solition moves according to the Landau--Lifshitz equation of motion, but as discussed in \cite{PMD}, \cite{KiePRD},
and \cite{CarKie}, its radiation-reaction term vanishes identically for the straight-line motion of a charged particle along a
constant applied electric field.
}
\begin{figure}[H]
  \includegraphics[width = 8truecm,scale=3]{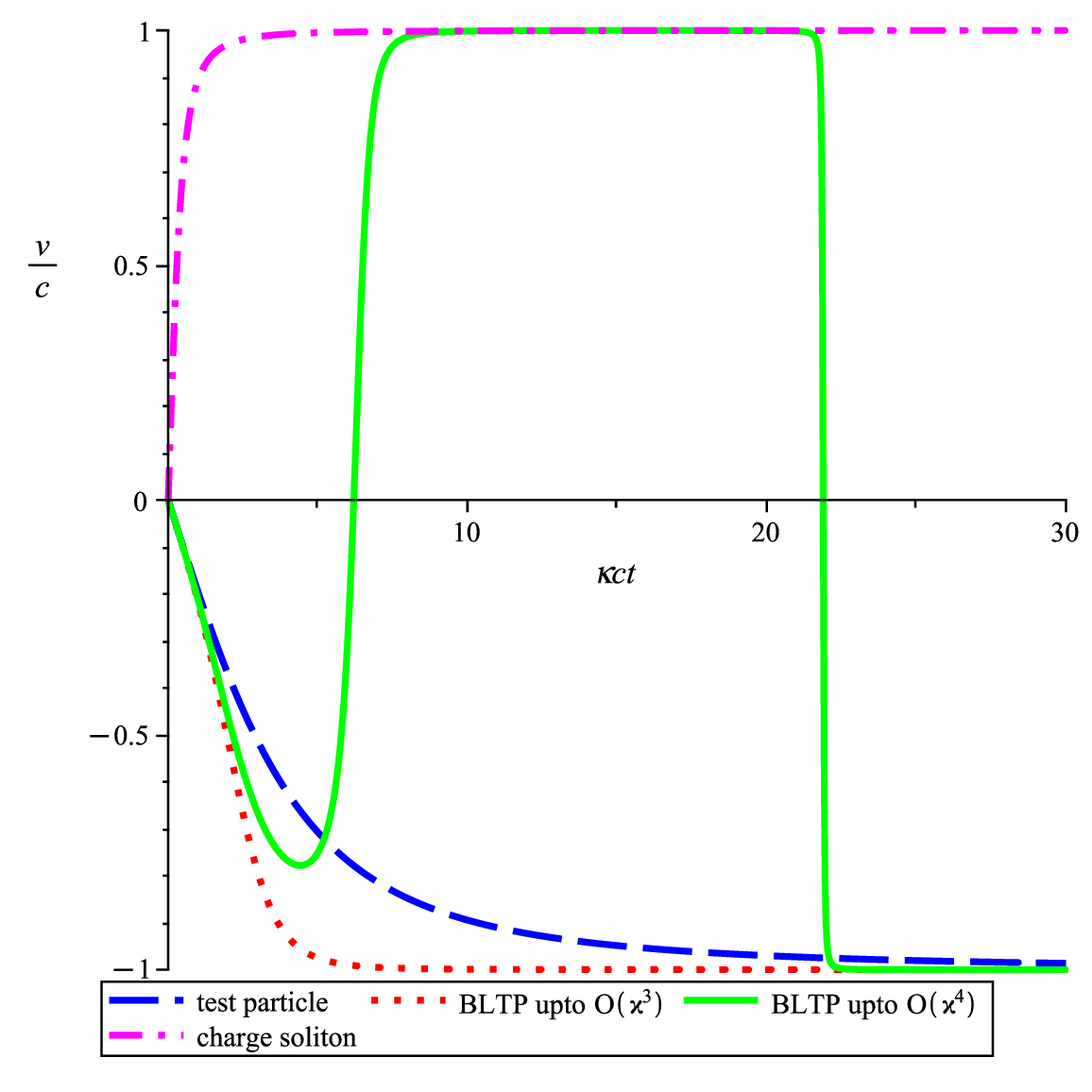} \vspace{-.3truecm}
\caption{
\footnotesize{The velocity of a point charge with negative bare mass $\mbare<0$, 
starting from rest in a constant applied electrostatic 
 field $\EV^{\mbox{\tiny{hom}}}=0.1e\varkappa^2$, vs. time, as per test particle theory (dashed curve), resp. BLTP electrodynamics
 with radiation-reaction included to $O(\varkappa^3)$ (dotted curve), resp. to
 $O(\varkappa^4)$ (continuous curve), when $\varkappa e^2/\mbare c^2 = -2$.
 Also shown is the radiation-free motion of a ``charge soliton,'' 
identical to charged test particle motion with effective mass $\mEL = \mbare + E^{\mbox{\tiny{field}}}/c^2 >0$,
though with $\mEL$ about 1000 times larger than it should, for optical purposes. }
}\vspace{-10pt}
\end{figure}

 It is interesting to see that in the $O(\varkappa^4)$ approximation to the motion in a BLTP vacuum, the early motion
``in the wrong direction''\footnote{\color{black}Incidentally, since it is the bare mass that supplies the inertia for the 
particle's motion in the initial phase, experimental studies of the initial phase can reveal the bare mass of the empirical 
electron \emph{in principle}.
 An intriguing question is whether such experimental studies are practically feasible.\vspace{-10pt}}
is soon self-corrected into the physically familiar motion in the right direction of the
dressed particle with positive effective mass.
 Unfortunately, after a while the $O(\varkappa^4)$ approximate motion switches back to the motion in the wrong direction, 
and in fact continues to switch back and forth after larger and larger intervals of steady motion. 
 This is the well-known signature of so-called ``over-stability,'' which usually means that a
perturbation of some equilibrium in some direction is counter-acted in the opposite direction, 
as in a stable situation, but in the over-stable case 
is in fact overcompensated, again and again, leading to a run-away in amplitude from the equilibrium situation.

 All this has to be swallowed with a grain of salt, because Fig.~4 is for $\frac12\varkappa e^2/\mbare c^2 = -1$,
so that approximate solutions at any finite $O(\varkappa^k)$ cannot be trusted to accurately reflect the actual motions
unless $\varkappa ct \ll 1$. 
 All the same, the study of the $O(\varkappa^4)$ approximation to the BLTP motion over longer times
 has the benefit of pointing out a dynamical possibility, namely the emergence of the empirical physical 
long-time motion after self-correcting the short-time motion{, which is} 
in the opposite direction of the long-time motion.
\vspace{-10pt}

%%%%%%%%%%%%%%%%%%%%%%%%%%%%%%%%%%%%%%%%%%%%%%%%%%%%%%%%%%%%
%%%%%%%%%%%%%%%%%%%%%%%%%%%%%%%%%%%%%%%%%%%%%%%%%%%%%%%%%%%%%%
%%%%%%%%%%%%%%%%%%%%%%%%%%%%%%%%%%%%%%%%%%%%%%%%%%%%%%%%%%%%%%%%%%%%
%%%%%%%%%%%%%%%%%%%%%%%%%%%%%%%%%%%%%%%%%%%%%%%%%%%%%%%%%%%%%%
     \section{Summary and Outlook}
\vspace{-5pt}
%%%%%%%%%%%%%%%%%%%%%%%%%%%%%%%%%%%%%%%%%%%%%%%%%%%%%%%%%%%%%%%%%%%%
%%%%%%%%%%%%%%%%%%%%%%%%%%%%%%%%%%%%%%%%%%%%%%%%%%%%%%%%%%%%%%
%%%%%%%%%%%%%%%%%%%%%%%%%%%%%%%%%%%%%%%%%%%%%%%%%%%%%%%%%%%%%%
%%%%%%%%%%%%%%%%%%%%%%%%%%%%%%%%%%%%%%%%%%%%%%%%%%%%%%%%%%%%%% 

In \cite{CarKie} it was shown that BLTP electrodynamics passes a litmus test that other models (in particular, 
the Landau--Lifshitz and Eliezer--Ford--O'Connell equations of motion) fail, namely that 
radiation-reaction features in the straight line motion of a point charge accelerated by a constant applied 
electric field in a BLTP vacuum.
 Although the results in \cite{CarKie} and those presented here are based on a ``small-$\varkappa$'' expansion 
of the BLTP force expression, for a \emph{proof-of-concept} demonstration  this is acceptable.

{The $O(\varkappa^3)$ 
results of \cite{CarKie} had raised a serious concern,} though: if the motions generated by the 
force with terms of $O(\varkappa^3)$ included are good approximations to the true BLTP motions over very long times as long as 
the dimensionless parameter $\varkappa e^2 / |\mbare| c^2 \ll 1$, then BLTP electrodynamics is presumably not a physically
reasonable model of classical electrodynamics.
 The cautious ``presumably'' is needed because physical viability of BLTP electrodynamics over relevantly long times 
has to hold for when $\varkappa e^2 / \mbare c^2 \approx - 2$.

{A negative bare mass, $\mbare<0$, of a point charge in 
a BLTP vacuum is suggested both by (a) general relativity, where the analogous problem of motion of point charges 
seems only to be well-definable when $\mbare<0$, due to the occurrence of black holes when $\mbare>0$ \cite{BKTZ}; 
and by (b) computation of the hydrogen spectrum in a BLTP vacuum \cite{CKP}, which demand such a large $\varkappa$ that the 
electrostatic self-energy of a point charge is about $10^4$ times larger than the empirical rest mass of an electron,
implying a huge negative bare mass to obtain the empirical value for  $\mEL = \mbare + E^{\mbox{\tiny{field}}}/c^2 >0$,  where
$E^{\mbox{\tiny{field}}}/c^2$ is the electrostatic energy of the charged particle at rest. 
 Since $E^{\mbox{\tiny{field}}}=\frac12\varkappa e^2$ in BLTP electrodynamics, this yields $\varkappa e^2/\mbare c^2 \approx -2$.}

 In this paper we have shown that BLTP electrodynamics with point charges, as defined in \cite{KiePRD}, remains 
a viable classical electrodynamical theory by demonstrating that the physically bizarre long-time motion of the 
``small $\varkappa$'' approximation truncated after $O(\varkappa^3)$ is not reproduced when the $O(\varkappa^4)$ 
terms are taken into account.
 This establishes that the $O(\varkappa^3)$ truncation cannot be trusted to produce good 
approximations to the actual solutions of BLTP electrodynamics over longer times even when the dimensionless 
parameter $\varkappa e^2 / |\mbare| c^2 \ll 1$.

 At short times the $O(\varkappa^4)$ motions and the $O(\varkappa^3)$ motions are very close, indicating convergence
of the ``small-$\varkappa$'' expansion for short times.
 This means that the ``small $\varkappa$'' expansion truncated after $O(\varkappa^3)$ is mathematically 
accurate for short times, i.e. when $\varkappa c t \ll 1$, and tentatively also for longer times $\varkappa ct > 1$
when the dimensionless parameter $\varkappa e^2 / |\mbare| c^2 \ll 1$, but not for arbitrarily long times 
even if the dimensionless parameter $\varkappa e^2 / |\mbare| c^2 \ll 1$.

 Good news for BLTP electrodynamics: it remains a viable classical physical theory, for the  unphysical
long-time behavior of its $O(\varkappa^3)$ solutions does not represent its true solutions.
 The situation gets more intriguing if we ask the question whether the  $O(\varkappa^4)$ solutions look physically
reasonable.

 For positive bare mass, $\mbare>0$, both the short time and the long-time behavior of the $O(\varkappa^4)$
 solutions looks physically (moderately) reasonable, 
with the velocity monotonically approaching the speed of light similar to the test particle motion.

For negative bare mass, $\mbare<0$, the short time behavior of the $O(\varkappa^4)$ solutions follows closely
the physically unfamiliar behavior of both the test particle motion and the $O(\varkappa^3)$ solutions, but eventually
departs strongly from these and for intermediate times switches over to the behavior of the charge soliton.
 Accelerated along a constant applied electric field a
 charge soliton carries out a test particle motion, though not of the bare charged
particle, but of the dressed particle with effective rest mass $\mEL = \mbare + E^{\mbox{\tiny{field}}}/c^2 >0$.
 The upshot is that the unfamiliar short time
behavior of a negative-bare-mass point charge in a BLTP vacuum may not be so unphysical after all, confined to 
some potentially very short transition time, after which the motion quickly resembles what is known from 
classical textbook theory,
though of course featuring radiation-reaction while textbook theory fails to do so.

Unfortunately, over sufficiently long times the $O(\varkappa^4)$ solution keeps switching back and forth between
a motion resembling the physically moderately reasonable behavior of the charge soliton (only ``moderately reasonable'' because 
the charge solition motion is radiation-reaction-free test particle motion in this setting), 
and something resembling the clearly unphysical long-time behavior of a test particle with negative bare mass.
 Thus, to establish the viability of BLTP electrodynamics in this problem one has to 
show that the unphysical back-and-forth of the long-time behavior of the $O(\varkappa^4)$ 
is absent from the true BLTP solutions.

In the small-$\varkappa$ regime  $\varkappa e^2/|\mbare|c^2\ll 1$ it may suffice to push the evaluation of the 
small-$\varkappa$ expansion further, certainly to order $\varkappa^5$, but perhaps further yet, and study whether
such oscillatory solutions disappear at higher order.

Of course, it is conceivable that the power series does not converge for arbitrarily long times even 
if $\varkappa e^2/|\mbare|c^2\ll 1$, and truncated at any order produce some kind of bizarre behavior over long enough times. 
 The ultimate answer can come from well-approximated BLTP solutions that do not rely on the small-$\varkappa$ expansion.

 In future research we hope to address the truly late-times regime, i.e. symbolically the ``large $\varkappa$'' regime
in both senses, for the dimensionless parameter $\varkappa e^2 / |\mbare| c^2 \approx 2$ and the dimensionless
time $\varkappa ct \gg 1$.
 This is a real challenge, for one cannot {Taylor-}expand around $\varkappa=\infty$.
 Recall that $\varkappa=\infty$ would throw us back to the ill-defined equations of Lorentz electrodynamics with
point charges.
 Moreover, it is not clear how far one could push to larger $t$ by
evaluating the small-$\varkappa$ expansion to many more orders in $\varkappa^n$, by making $n$ large enough.
{Intriguingly, though, the $O(\varkappa^3)$ and  $O(\varkappa^4)$ contributions to the self-force are linear.
 This raises the question whether all higher-order contribution are linear, too, and
in itself encourages pushing the small $\varkappa$ expansion to higher order.}
\smallskip

\noindent 
\textbf{Acknowledgement:} The authors are grateful to Claus L\"ammerzahl, Volker Perlick,
and Alessandro Spallicci, the organizers of the workshop ``Developments, Problems, and Extensions of Electrodynamics,''
18--23 Aug. 2024, in Bad Honnef / Germany, where the authors met and started their collaboration on the
topic of this paper. 
 The authors also extend their thanks to Lilit Sargsyan for spotting a couple of minus sign mistakes in the appendix of an earlier preprint version of this paper; happily, the main conclusions of this paper were not affected.

\newpage
\noindent
{\Large\textbf{APPENDIX}}
\begin{appendix}

\section{The self-force on a charge in a BLTP vacuum}

 The self-field force can be evaluated using retarded spherical coordinates $(r,\vartheta,\varphi)$ to carry out the
$\drm^3{s}$ integrations over the ball ${B_{ct}(\qV_0)}$, after which one can differentiate w.r.t. $t$.
 For the problem of straight line motion of a charge starting from rest at the origin, 
this yields
\begin{alignat}{2}\label{eq:selfFexpl}
\hspace{-20pt}
\fV^{\mbox{\tiny{self}}}[\qV,\vV;{\color{red}\aV}](t)
&=  \frac{\ee^2} {4\pi } \biggl[ \biggr.
 - {\mathbf{Z}}_{\bxi}^{[2]}(t,t) 
\\ \notag
& \qquad\quad  -\!\!\! \;{\textstyle\sum\limits_{0\leq k\leq 1}}\! c^{2-k}(2-k)\!\!
\displaystyle  \int_0^{t}\!  
{\mathbf{Z}}_{\bxi}^{[k]}\big(t,t^{\mathrm{r}}\big)
(t- t^{\mbox{\tiny{r}}})^{1-k} \drm{t^{\mbox{\tiny{r}}}} 
\\ \notag
& \qquad\quad -\!\!\!  \;{\textstyle\sum\limits_{0\leq k\leq 2}}\! c^{2-k}\!
\displaystyle  \int_0^{t}\!  
\tpddt{\mathbf{Z}}_{\bxi}^{[k]}\big(t,t^{\mathrm{r}}\big)
(t- t^{\mathrm{r}})^{2-k} \drm{t^{\mathrm{r}}}  \biggl. \biggr].
\end{alignat}
 Here, $\bxi(t) \equiv (\qV,\vV,{\color{red}\aV})(t)$  and 
${\mathbf{Z}}_{\bxi}^{[2]}(t,t) :=\lim_{t^{\mathrm{r}}\to t}
{\mathbf{Z}}_{\bxi}^{[2]}\big(t,t^{\mathrm{r}}\big)$,
where for $k\in\{0,1,2\}$, 
\begin{alignat}{1}\label{eq:Zdef}
{\mathbf{Z}}_{\bxi}^{[k]}\big(t,t^{\mathrm{r}}\big) = 
 \displaystyle  \int_0^{2\pi}\!\! \int_0^{\pi}\!
\left(1-\beta(t^\mathrm{r})\cos\vartheta\right)
 \boldsymbol{\pi}_{\bxi}^{[k]}\big(t,\qV(t^\mathrm{r}) + c(t-t^\mathrm{r})\nV 
\big) 
\sin\vartheta \drm{\vartheta}\drm{\varphi}\,,
 \end{alignat}
with $\beta(t) |\EV^{\mbox{\tiny{hom}}}| \equiv\bsym{\beta}(t)\cdot\EV^{\mbox{\tiny{hom}}}$,
where $\bsym{\beta}(t) :=\frac1c\vV(t)$ in standard notation; moreover, the unit vector
$\nV =\left(\sin\vartheta \cos\varphi,\;\sin\vartheta \sin\varphi ,\; \cos\vartheta \right)$
is normal to the retarded sphere of radius $r=c(t-t^\mathrm{r})$, with $\vartheta$ counted 
from the $\EV^{\mbox{\tiny{hom}}}$ direction and $\varphi$ from an arbitrary axis $\perp\EV^{\mbox{\tiny{hom}}}$.

 Furthermore, the $\boldsymbol{\pi}_{\bxi}^{[k]}(t,\sV)$ with $k\in\{0,1,2\}$ and $\sV\neq\qV$ are defined as follows.
 We set 
\begin{alignat}{1}
\mathrm{K}_{\bxi}(t',t,\sV) & := \label{rmK}
\tfrac{J_1\!\bigl(\varkappa\sqrt{c^2(t-t')^2-|\sV-\qV(t')|^2 }\bigr)}{\sqrt{c^2(t-t')^2-|\sV-\qV(t')|^2}^{\phantom{n}}},\\
\mathbf{K}_{\bxi}(t',t,\sV) & := \label{bfK}
\tfrac{J_2\!\bigl(\varkappa\sqrt{c^2(t-t')^2-|\sV-\qV(t')|^2 }\bigr)}{{c^2(t-t')^2-|\sV-\qV(t')|^2}^{\phantom{n}} }
 \left(\sV-\qV(t')- \vV(t')(t-t')\right), 
\end{alignat}
and with ${\bxi}^\circ$ shorthand for trivial map $t\mapsto(\boldsymbol{0},\boldsymbol{0},\boldsymbol{0})$, 
we note that
\begin{equation}
\label{eq:7}
\mathrm{K}_{\bxi^\circ}(t',t,\sV) 
=\frac{J_{1}(\varkappa \sqrt{c^{2}(t-t')^{2}-|\sV|^{2}})}{ \sqrt{c^{2}(t-t')^{2}-|\sV|^{2}}},
\end{equation}
\begin{equation}
\label{eq:8}
 \mathbf{K}_{\bxi^\circ}(t',t,\sV)
=\frac{J_{2}(\varkappa \sqrt{c^{2}(t-t')^{2}-|\sV|^{2}})}{ c^{2}(t-t')^{2}-|\sV|^{2}}\sV.
\end{equation}
 We also note that 
\begin{alignat}{1}
\label{bfKxiNULL}
\int_{0}^{t^\mathrm{ret}_{\bxi^\circ}(t,\sV)} \!\!\!\!
 \mathbf{K}_{\bxi^\circ}(t',t,\sV)c\drm{t'} 
=
\int_0^{ct-|\sV|}\frac{J_2(\varkappa\sqrt{c^2(t-t')^2-|\sV|^2})}{c^2(t-t')^2-|\sV|^2}\sV\,\drm(ct'),
\end{alignat}
\begin{alignat}{1}
\label{rmKxiNULL}
 \int_{0}^{t^\mathrm{ret}_{\bxi^\circ}(t,\sV)}  \!\!\!\!\mathrm{K}_{\bxi^\circ}(t',t,\sV) c\drm{t'}
=
\int_0^{ct-|\sV|}\frac{J_1(\varkappa\sqrt{c^2(t-t')^2-|\sV|^2})}{\sqrt{c^2(t-t')^2-|\sV|^2}}\,\drm(ct').
\end{alignat}
 We use $\big|^\mathrm{ret}$ to indicate that $\qV(\tilde{t})$, $\vV(\tilde{t})$, $\bbeta(\tilde{t})$,
${\color{red}\aV}(\tilde{t})$ are evaluated at~$\tilde{t} = {t^\mathrm{ret}_{\bxi}}(t,\sV)$,
\textit{not} at ${t^\mathrm{ret}_{\boldsymbol{\xi^\circ}}}(t,\sV)$.
   Then, with $\nV(\qV,\sV)$ denoting the unit vector from $\qV$ to $\sV$, we have
\begin{alignat}{1}
\label{pi2}
\boldsymbol{\pi}_{\bxi}^{[2]}(t,\sV) = & - \varkappa^2
\left[\textstyle\frac{1}{\bigl({1- {\bsym{\beta}}\cdot\nV(\qV,\sV)}\bigr)^{\!2} }\bsym{\beta}
 - \Big[\!{1-\beta^2}\!\Big]
\frac{ \left({\nV(\qV,\sV)}-\bsym{\beta}\right) \crprd \left(\bsym{\beta}\crprd \nV(\qV,\sV)\right) }{
      \bigl({1-\bsym{\beta}\cdot\nV(\qV,\sV)}\bigr)^4 }
\right]^{\mathrm{ret}} 
\\ \notag
&  +\varkappa^2 \left[\Big[\!{1-\beta^2}\!\Big] {\textstyle{
\frac{ \bbeta\crprd {\nV(\qV,\sV)}_{\phantom{!\!}} }{
      \bigl({1-\bbeta\cdot\nV(\qV,\sV)}\bigr)^{\!3} }
}}\right]^{\mathrm{ret}} \crprd\int_{0}^{t^\mathrm{ret}_{\bxi}(t,\sV)} \!\!\!\!
\mathbf{K}_{\bxi}(t',t,\sV)c\drm{t'} \\
\notag
&  - \varkappa^2\left[\Big[\!{1-\beta^2}\!\Big] {\textstyle{
\frac{ \bbeta \crprd {\nV(\qV,\sV)}_{\phantom{!\!}} }{
      \bigl({1-\bbeta\cdot\nV(\qV,\sV)}\bigr)^{\!3} }
}}\right]^{\mathrm{ret}} \crprd\left(
 \tfrac{1-(1+\varkappa|\sV|)e^{-\varkappa|\sV|}}{\varkappa^2|\sV|^2}-\tfrac12\right)\tfrac{\sV}{|\sV|}
\\
\notag  
&  -\varkappa^2 \left[\Big[\!{1-\beta^2}\!\Big]{\textstyle{
\frac{\bbeta\crprd  {\nV(\qV,\sV)}_{\phantom{!\!}} }{
      \bigl({1-\bbeta\cdot\nV(\qV,\sV)}\bigr)^{\!3} }
}}\right]^{\mathrm{ret}} \crprd \int_{0}^{t^\mathrm{ret}_{\bxi^\circ}(t,\sV)} \!\!\!\!
\mathbf{K}_{\bxi^\circ}(t',t,\sV)c\drm{t'} 
\\
\notag
& -  \varkappa^2\left[\Big[\!{1-\beta^2}\!\Big]
{\textstyle{
\frac{ {\nV(\qV,\sV)}_{\phantom{!\!}}-\bbeta }{
      \bigl({1-\bbeta\cdot\nV(\qV,\sV)}\bigr)^{\!3} }
}}\right]^{\mathrm{ret}}
\crprd \!
\int_{0}^{t^\mathrm{ret}_{\bxi}(t,\sV)} \!\!\!\!
{\bbeta(t')}\crprd \mathbf{K}_{\bxi}(t',t,\sV)c\drm{t'} 
\end{alignat}
and
\begin{alignat}{1}
\label{pi1}
 \boldsymbol{\pi}_{\bxi}^{[1]}(t,\sV) = 
& - \varkappa^2 
\left[
{\textstyle{
{{\nV(\qV,\sV)}\frac{\left({\nV(\qV,\sV)}\crprd [{ 
{\nV(\qV,\sV)} 
\crprd {\color{red}\aV} }]\right)\cdot\bbeta}{
       c^2 \bigl({1-\bbeta\cdot\nV(\qV,\sV)}\bigr)^{\!4} }
    } +
{\nV(\qV,\sV)} \crprd\frac{ {\nV(\qV,\sV)} 
\crprd {\color{red}\aV} }{
      2c^2 \bigl({1-\bbeta\cdot\nV(\qV,\sV)}\bigr)^{\!3} }
}}\right]^{\mathrm{ret}}\\ \notag
&- \varkappa^2\left[
{\textstyle{
{\nV(\qV,\sV)}\crprd\frac{ {\nV(\qV,\sV)} 
\crprd {\color{red}\aV} }{
      c^2 \bigl({1-\bbeta\cdot\nV(\qV,\sV)}\bigr)^{\!3} }
}}\right]^{\mathrm{ret}} \!\!
\crprd \!
\int_{0}^{t^\mathrm{ret}_{\bxi}(t,\sV)}\!\!\!\!
{\bbeta(t')}\crprd \mathbf{K}_{\bxi}(t',t,\sV)c\drm{t'} 
\\ \notag
& + \varkappa^2\left[\nV(\qV,\sV)\crprd \biggl[{\textstyle{\nV(\qV,\sV)\crprd 
\frac{{\nV(\qV,\sV)}_{\phantom{!\!}} 
\crprd{\color{red}\aV} }{
      c^2\bigl({1-\bbeta\cdot\nV(\qV,\sV)}\bigr)^{\!3} }
}}\biggr]\right]^{\mathrm{ret}} \!\!\! 
\crprd\! \int_{0}^{t^\mathrm{ret}_{\bxi}(t,\sV)} \!\!\!\!
\mathbf{K}_{\bxi}(t',t,\sV)c\drm{t'} 
\\ \notag
& - \varkappa^2\left[\nV(\qV,\sV)\crprd \biggl[{\textstyle{\nV(\qV,\sV)\crprd 
\frac{{\nV(\qV,\sV)}_{\phantom{!\!}} 
\crprd{\color{red}\aV} }{
      c^2\bigl({1-\bbeta\cdot\nV(\qV,\sV)}\bigr)^{\!3} }
}}\biggr]\right]^{\mathrm{ret}} \!\!\! 
\crprd\! \left(
 \tfrac{1-(1+\varkappa|\sV|)e^{-\varkappa|\sV|}}{\varkappa^2|\sV|^2}-\tfrac12\right)\tfrac{\sV}{|\sV|}\
\\ \notag
& - \varkappa^2\left[\nV(\qV,\sV)\crprd \biggl[{\textstyle{\nV(\qV,\sV)\crprd 
\frac{{\nV(\qV,\sV)}_{\phantom{!\!}} 
\crprd{\color{red}\aV} }{
      c^2\bigl({1-\bbeta\cdot\nV(\qV,\sV)}\bigr)^{\!3} }
}}\biggr]\right]^{\mathrm{ret}} \!\!\! 
\crprd\! \int_{0}^{t^\mathrm{ret}_{\bxi^\circ}(t,\sV)}\!\!\!\!\mathbf{K}_{\bxi^\circ}(t',t,\sV)c\drm{t'}
 \\ \notag
& +\varkappa^3 
 \left[\textstyle\frac{1}{{1-\bbeta\cdot\nV(\qV,\sV)} }\right]^{\mathrm{ret}} 
 \int_{0}^{t^\mathrm{ret}_{\bxi}(t,\sV)}\!\!\!\!
 \mathrm{K}_{\bxi}(t',t,\sV)\left[{\bbeta}({t^\mathrm{ret}_{\bxi}(t,\sV)})+{\bbeta}(t')\right]c\drm{t'}\,,
\\ \notag
& +\varkappa^2 
 \left[\textstyle\frac{1}{{1-\bbeta\cdot\nV(\qV,\sV)} }\right]^{\mathrm{ret}} 
\tfrac{1-e^{-\varkappa|\sV|}}{|\sV|}\bbeta({t^\mathrm{ret}_{\bxi}(t,\sV)})
\\ \notag
& -\varkappa^3 
 \left[\textstyle\frac{1}{{1-\bbeta\cdot\nV(\qV,\sV)} }\right]^{\mathrm{ret}} 
 \int_{0}^{t^\mathrm{ret}_{\bxi^\circ}(t,\sV)}  \!\!\!\!\mathrm{K}_{\bxi^\circ}(t',t,\sV) c\drm{t'}
\bbeta({t^\mathrm{ret}_{\bxi}(t,\sV)}),
 \end{alignat}\vspace{-.5truecm}
and
\begin{alignat}{1}
\label{pi0}
 \boldsymbol{\pi}_{\bxi}^{[0]}(t,\sV) =
& - \varkappa^4 \frac14\left[
{\textstyle{
\frac{\left({\nV(\qV,\sV)} -\bbeta\right)\crprd\left(\bbeta\crprd {\nV(\qV,\sV)} \right)}{
      \bigl({1-\bbeta\cdot\nV(\qV,\sV)}\bigr)^{\!2} }
           }}\right]^{\mathrm{ret}}\\ \notag
&+ \varkappa^4\frac12\left[
{\textstyle{
\frac{ {\nV(\qV,\sV)}
-\bbeta}{ {1-\bbeta\cdot\nV(\qV,\sV)} }
             }}\right]^{\mathrm{ret}} 
\crprd \!
\int_{0}^{t^\mathrm{ret}_{\bxi}(t,\sV)}\!\!\!\!
{\bbeta(t')}\crprd \mathbf{K}_{\bxi}(t',t,\sV)c\drm{t'} 
\\ \notag
& - \varkappa^4\frac12\left[{\textstyle{\frac{ \bbeta\crprd {\nV(\qV,\sV)} }{
      1-\bbeta\cdot\nV(\qV,\sV)} }}\right]^{\mathrm{ret}} 
\crprd 
\int_{0}^{t^\mathrm{ret}_{\bxi}(t,\sV)}\!\!\!\!
\mathbf{K}_{\bxi}(t',t,\sV)c\drm{t'}
\\  \notag
&
 +\varkappa^4\frac12\left[{\textstyle{\frac{ \bbeta\crprd {\nV(\qV,\sV)} }{
      1-\bbeta\cdot\nV(\qV,\sV)} }}\right]^{\mathrm{ret}} 
\crprd 
\left(
 \tfrac{1-(1+\varkappa|\sV|)e^{-\varkappa|\sV|}}{\varkappa^2|\sV|^2}-\tfrac12\right)\tfrac{\sV}{|\sV|}
\\ \notag
&+\varkappa^4\frac12\left[{\textstyle{\frac{  \bbeta\crprd {\nV(\qV,\sV)} }{
      1-\bbeta\cdot\nV(\qV,\sV)} }}\right]^{\mathrm{ret}} 
\crprd \int_{0}^{t^\mathrm{ret}_{\bxi^\circ}(t,\sV)} \!\!\!\! \mathbf{K}_{\bxi^\circ}(t',t,\sV)c\drm{t'} 
\\ \notag
& - \varkappa^4 \int_{0}^{t^\mathrm{ret}_{\bxi}(t,\sV)} \!\!\!\!
\mathbf{K}_{\bxi}(t',t,\sV)c\drm{t'} \crprd \int_{0}^{t^\mathrm{ret}_{\bxi}(t,\sV)} \!\!\!\!
{\bbeta(t')}\crprd \mathbf{K}_{\bxi}(t',t,\sV)c\drm{t'} 
\\ \notag
& +\varkappa^4 \left(
 \tfrac{1-(1+\varkappa|\sV|)e^{-\varkappa|\sV|}}{\varkappa^2|\sV|^2}-\tfrac12\right)\tfrac{\sV}{|\sV|}\crprd 
 \int_{0}^{t^\mathrm{ret}_{\bxi}(t,\sV)} \!\!\!\!
{\bbeta(t')}\crprd\mathbf{K}_{\bxi}(t',t,\sV)c\drm{t'} 
\\ \notag
&+\varkappa^4
\int_{0}^{t^\mathrm{ret}_{\bxi^\circ}(t,\sV)} \!\!\!\! \mathbf{K}_{\bxi^\circ}(t',t,\sV)c\drm{t'} 
\crprd \int_{0}^{t^\mathrm{ret}_{\bxi}(t,\sV)} \!\!\!\!
{\bbeta(t')}\crprd \mathbf{K}_{\bxi}(t',t,\sV)c\drm{t'} 
\\ \notag
& - \varkappa^4 \int_{0}^{t^\mathrm{ret}_{\bxi}(t,\sV)} \!\!\!\! \mathrm{K}_{\bxi}(t',t,\sV)c\drm{t'} 
 \int_{0}^{t^\mathrm{ret}_{\bxi}(t,\sV)}  \!\!\!\!\mathrm{K}_{\bxi}(t',t,\sV) {\bbeta}(t')c\drm{t'}\,,
\\
\notag
& - \varkappa^3 \tfrac{1-e^{-\varkappa|\sV|}}{|\sV|}
 \int_{0}^{t^\mathrm{ret}_{\bxi}(t,\sV)}  \!\!\!\!\mathrm{K}_{\bxi}(t',t,\sV) {\bbeta}(t')c\drm{t'}\,,
\\ \notag
& + \varkappa^4
 \int_{0}^{t^\mathrm{ret}_{\bxi^\circ}(t,\sV)}  \!\!\!\!\mathrm{K}_{\bxi^\circ}(t',t,\sV) c\drm{t'}
 \int_{0}^{t^\mathrm{ret}_{\bxi}(t,\sV)}  \!\!\!\!\mathrm{K}_{\bxi}(t',t,\sV) {\bbeta}(t')c\drm{t'}\,.
\end{alignat}

\section{Computing the self-force at $O(\varkappa^4)$}
The goal of Appendix B is to explicitly calculate the 4th order contribution to the self-force in the small-$\varkappa$ regime. 
 To compute the self-force we need to expand
the vectors $\bld{Z}_{\bxi}^{[k]}$ and $\pt{t}\bld{Z}_{\bxi}^{[k]}$, with $k\in\{0,1,2\}$, in 
powers of $\varkappa$.
 For this in turn we begin by expanding their constituent terms $\bsym{\pi}_{\bxi}^{[k]}$ about $\varkappa=0$ 
and ignore all terms of higher and lower order than 4.

 Writing $\bsym{\pi}_{\bxi}^{[k]}=\sum_{i}\bsym{\pi}_{\bxi}^{[k],i}$, where $i$ enumerates the 
lines at r.h.s.(\ref{pi2}), resp. r.h.s.(\ref{pi1}), resp. r.h.s.(\ref{pi0}), we find that for the 
non-derivative terms the following $\bsym{\pi}_{\bxi}^{[k],i}$ terms will make non-vanishing contributions at 4th order:
\begin{align}
\label{eq:4a}
&\bsym{\pi}_{\bxi}^{[0],1}(t,\sV)=
-\frac{\varkappa^{4}}{4}\lt[\frac{(\nh-\bsym{\beta})\crprd (\bsym{\beta}\crprd \nh)}{(1-\bsym{\beta}\cdot \nh)^{2}}\rt]\re
\\
\label{eq:4b}
&\bsym{\pi}_{\bxi}^{[1],2}(t,\sV)=
-\varkappa^{2}\lt[\nh\crprd\frac{(\nh\crprd {\color{red}\aV})}{c^{2}(1-\bsym{\beta}\cdot \nh )^{3}}\rt]\re \crprd 
\int_{0}^{t\re_{\bxi} } \bbeta(t') \crprd \bld{K}_{\bxi}(t',t,\sV)c\drm{t'}
\\
\label{eq:4c}
&\bsym{\pi}_{\bxi}^{[1],3}(t,\sV)=
\varkappa^{2}\lt[\nh\crprd\lt[\nh\crprd\frac{(\nh\crprd {\color{red}\aV})}{c^{2}(1-\bsym{\beta}\cdot \nh )^{3}}\rt]\rt]\re \crprd 
\int_{0}^{t\re_{\bxi} } \bld{K}_{\bxi}(t',t,\sV)c\drm{t'}
\end{align}
\begin{align}
\label{eq:4d}
&\bsym{\pi}_{\bxi}^{[1],4}(t,\sV)=
-\varkappa^{2}\lt[\nh\crprd\!\lt[\nh\crprd\frac{(\nh\crprd {\color{red}\aV})}{c^{2}(1-\bsym{\beta}\cdot \nh )^{3}}\rt]\rt]\re
\!\!\crprd\! 
\lt(\!\frac{1-(1+\varkappa|\sV|)e^{-\varkappa |\sV|}}{\varkappa^{2}|\sV|^{2}}-\frac{1}{2}\rt)\!\frac{\sV}{|\sV|}
\\
\label{eq:4e}
&\bsym{\pi}_{\bxi}^{[1],5}(t,\sV)=
-\varkappa^{2}\lt[\nh\crprd\lt[\nh\crprd\frac{(\nh\crprd {\color{red}\aV})}{c^{2}(1-\bsym{\beta}\cdot \nh )^{3}}\rt]\rt]\re \crprd 
\int_{0}^{t_{\bxi^\circ}\re } \bld{K}_{\bxi^\circ}(t',t,\sV)c\drm{t'}
\\
\label{eq:4f}
&\bsym{\pi}_{\bxi}^{[1],6}(t,\sV)=
\varkappa^{3}\lt[\frac{1}{(1-\bsym{\beta}\cdot \nh)}\rt]\re
\int_{0}^{t\re_{\bxi} }
\mathrm{K}_{\bxi}(t',t,\sV)[\bbeta\re+\bbeta(t)]c\drm{t'}
\\
\label{eq:4g}
&\bsym{\pi}_{\bxi}^{[1],7}(t,\sV)=
\varkappa^{2}\frac{1-e^{-\varkappa|\sV|}}{|\sV|}\lt[\frac{\bsym{\beta}}{(1-\bsym{\beta}\cdot \nh)}\rt]\re
\\
\label{eq:4h}
&\bsym{\pi}_{\bxi}^{[1],8}(t,\sV)=
-\varkappa^{3}\lt[\frac{\bsym{\beta}}{(1-\bsym{\beta}\cdot \nh)}\rt]\re
\int_{0}^{t_{\bxi^\circ}\re }\mathrm{K}_{\bxi^\circ}(t',t,\sV)c\drm{t'}
\\
\label{eq:4i}
&\bsym{\pi}_{\bxi}^{[2],2}(t,\sV)=
\varkappa^{2}\lt[(1-\beta^{2})\frac{\bsym{\beta}\crprd \nh}{(1-\bsym{\beta}\cdot \nh)^{3}}\rt]\re \crprd 
\int_{0}^{t\re_{\bxi} }  \bld{K}_{\bxi}(t',t,\sV)c\drm{t'}
\\
\label{eq:4j}
&\bsym{\pi}_{\bxi}^{[2],3}(t,\sV)=
-\varkappa^{2}\lt[(1-\beta^{2})\frac{\bsym{\beta}\crprd \nh}{(1-\bsym{\beta}\cdot \nh)^{3}}\rt]\re \crprd \lt(\frac{1-(1+\varkappa|\sV|)e^{-\varkappa |\sV|}}{\varkappa^{2}|\sV|^{2}}-\frac{1}{2}\rt)\frac{\sV}{|\sV|}
\\
\label{eq:4k}
&\bsym{\pi}_{\bxi}^{[2],4}(t,\sV)=
-\varkappa^{2}\lt[(1-\beta^{2})\frac{\bsym{\beta}\crprd \nh}{(1-\bsym{\beta}\cdot \nh)^{3}}\rt]\re \crprd 
\int_{0}^{t_{\bxi^\circ}\re }\bld{K}_{\bxi^\circ}(t',t,\sV)c\drm{t'} 
\\
\label{eq:4l}
&\bsym{\pi}_{\bxi}^{[2],5}(t,\sV)=
-\varkappa^{2}\lt[(1-\beta^{2})\frac{ \nh-\bsym{\beta}}{(1-\bsym{\beta}\cdot \nh)^{3}}\rt]\re \crprd \int_{0}^{t\re }
\bbeta(t') \crprd \bld{K}_{\bxi}(t',t,\sV)c\drm{t'} 
\end{align}
where the variables denoted by the ``ret'' suffix are evaluated at the retarded time $t\re$ 
which is defined by $c(t-t\re)=|\sV-\qV(t\re)|$, e.g. $\qV\re=\qV(t\re)$. 
 Here generically the vector $\nh$ is tacitly a vector function $\nh(\qV,\sV)$. 
The retarded time denoted $t\re_{\bxi^\circ}$ for the auxiliary (trivial) trajectory
is simply defined by $t\re_{0}(t,\sV)=t-|\sV|/c$. 

We next begin expanding the \req{4a}--\req{4l} terms to 4th order in $\varkappa$.

\subsection{Identifying all order $\varkappa^4$ terms}
To extract the 4th order contributions from the $\boldsymbol{\pi}_{\bxi}^{[k],i}$ 
we must expand the pertinent Bessel function kernels, respectively the exponentials 
$e^{-\varkappa|\sV|}$ and identify those terms that, when multiplied by the pertinent pre-factors $\varkappa^j$, with
$j\in\{2,3\}$, in \req{4a}--\req{4l}, yield a total factor of  $\varkappa^{4}$. 

 We begin with the exponential 
\begin{equation}
\label{eq:11}
e^{-\varkappa |\sV|} =
 1-\varkappa |\sV|+\frac{\varkappa^{2} |\sV|^{2}}{2}-\frac{\varkappa^{3} |\sV|^{3}}{6}+\frac{\varkappa^{4} |\sV|^{4}}{24} +
 O(\varkappa^5|\sV|^5).
\end{equation}
With the help of (\ref{eq:11}) we find
\begin{equation}
\label{eq:13}
\frac{1-e^{-\varkappa|\sV|}}{\varkappa|\sV|} = 1 - \frac{\varkappa|\sV|}{2} + O(\varkappa^2|\sV|^2)
\end{equation}
and
\begin{align}
\label{eq:12}
\frac{1-(1+\varkappa|\sV|)e^{-\varkappa |\sV|}}{\varkappa^{2}|\sV|^{2}}-\frac{1}{2}
 = 
-\frac{\varkappa|\sV|}{3}+\frac{\varkappa^{2}|\sV|^{2}}{8} + O(\varkappa^3|\sV|^3)
\end{align}
 Inserting \req{13} in \req{4g} and identifying the fourth order term $\bsym{\pi}_{\bxi,(4)}^{[1],7}$, we find
\begin{align}
\label{eq:pi174}
\bsym{\pi}_{\bxi,(4)}^{[1],7}(t,\sV)
=
-\frac12\varkappa^{4}|\sV|\lt[\frac{\bsym{\beta}}{1-\bsym{\beta}\cdot \nh}\rt]\re
\end{align}
 Inserting \req{12} in \req{4d} and \req{4j}, then identifying the fourth order terms 
$\bsym{\pi}_{\bxi,(4)}^{[1],4}$, respectively $\bsym{\pi}_{\bxi,(4)}^{[2],3}$, we find
\begin{align}
\label{eq:pi144}
\bsym{\pi}_{\bxi,(4)}^{[1],4} (t,\sV)
&=
-\varkappa^{4}\frac{|\sV|}{8c^2}
\lt[\frac{({\color{red}\aV}\cdot \sV)\nh-(\nh\cdot \sV){\color{red}\aV}}
{(1-\bsym{\beta}\cdot \nh )^{3}}\rt]\re,
\\
\label{eq:pi234}
\bsym{\pi}_{\bxi,(4)}^{[2],3}(t,\sV)
&=
{-}
\varkappa^{4}\frac{|\sV|}{8}
\lt[\frac{(1-\beta^{2})\big((\sV\cdot \bsym{\beta})\nh-(\nh \cdot \sV)\bsym{\beta}\big)}{(1-\bsym{\beta}\cdot \nh )^{3}}\rt]\re.
\end{align}

 Coming next to the Bessel function kernels of the first kind $J_{n}$, their 
expansions read
\begin{equation}
\label{eq:9}
J_{1}(x) = \frac{x}{2}\lt(1-\frac{x^{2}}{8}+\frac{x^{4}}{192} + O(x^6)\rt),
\end{equation}
\begin{equation}
\label{eq:10}
J_{2}(x)=\frac{x^{2}}{8}\lt(1-\frac{x^{2}}{12}+\frac{x^{4}}{384}+ O(x^6)\rt).
\end{equation}
 Recalling that the terms containing the  Bessel function kernels
already have factors of $\varkappa^{2}$ or $\varkappa^{3}$, 
to arrive at the $O(\varkappa^4)$ contributions to the $\bsym{\pi}$s 
the leading order of the Bessel expansions (\ref{eq:9}), (\ref{eq:10}) is all that is required.
 Note that the integrals over $\mathrm{K}_{\bxi}$ and $\bld{K}_{\bxi}$ in 
(\ref{eq:4b}), (\ref{eq:4c}), (\ref{eq:4e}), (\ref{eq:4f}), (\ref{eq:4h}), (\ref{eq:4i}), (\ref{eq:4k}), (\ref{eq:4l})
 are rendered integrable when the Bessel functions are expanded to 0th order of(\ref{eq:10}), (\ref{eq:11}). 
 To this end we evaluate as follows
\begin{align}
\begin{split}
\label{eq:14}
\int_{0}^{t\re_{\bxi} } \bbeta(t') \crprd \bld{K}_{\bxi}(t',t,\sV)c\drm{t'}&
= \frac{\varkappa^{2}}{8} \int_{0}^{t\re_{\bxi} } \vV (t')\crprd \lt[\sV-\qV(t')-\vV(t')(t-t')\rt]\drm{t'} +O(\varkappa^4)
\\
&
=\frac{\varkappa^{2}}{8}\int_{0}^{t\re_{\bxi} } \vV(t') \drm{t'}\crprd \sV +O(\varkappa^4)
\\
&
=\frac{\varkappa^{2}}{8}\qV(t\re_{\bxi}) \crprd \sV +O(\varkappa^4),
\end{split}
\end{align}
where we have utilised the fact that $\qV(t) \ || \ \vV(t') \ ||\  \EV^{\mbox{\tiny{hom}}}$ 
for any choices of $t$ and $t'$ (i.e. the charge travels in a straight line).

 Likewise the other integrals that appear are easy enough to evaluate; thus
\begin{align}
\begin{split}
\label{eq:15}
\int_{0}^{t\re_{\bxio} }\bld{K}_{\bxio}(t',t,\sV)c\drm{t'}
 & =\frac{\varkappa^{2}}{8} \sV \int_{0}^{t\re_{\bxio} }c\drm{t'} +O(\varkappa^4)
\\
&
=\frac{\varkappa^{2}}{8}\sV\lt(ct-|\sV|\rt)+O(\varkappa^4)
 \end{split}
\end{align}
\begin{align}
\begin{split}
\label{eq:16}
\int_{0}^{t\re_{\bxio} } \mathrm{K}_{\bxio}(t',t,\sV)c\drm{t'} &=\frac{\varkappa}{2} \int_{0}^{t\re_{\bxio}}c\drm{t'} +O(\varkappa^3)
\\
&
=\frac{\varkappa}{2}\lt(ct-|\sV|\rt) +O(\varkappa^3)
\end{split}
\end{align}
\begin{align}
\begin{split}
\label{eq:17}
\int_{0}^{t\re_{\bxi} } \mathrm{K}_{\bxi}(t',t,\sV)[\bbeta\re+\bbeta(t')]c\drm{t'}& = 
\frac{\varkappa}{2}\int_{0}^{t\re_{\bxi} }[\vV(t\re_{\bxi})+\vV(t')]\drm{t'}+O(\varkappa^3)
\\
&=\frac{\varkappa}{2}\lt[\vV(t\re_{\bxi}) t\re_{\bxi}+\qV(t\re_{\bxi})\rt] +O(\varkappa^3)
\end{split}
\end{align}
\begin{align}
\begin{split}
\label{eq:18}
\int_{0}^{t\re_{\bxi} } \bld{K}_{\bxi}(t',t,\sV)c\drm{t'} & =
\frac{\varkappa^{2}}{8}\int_{0}^{t\re_{\bxi}}(\sV-\qV(t')-\vV(t')(t-t'))c\drm{t'} +O(\varkappa^4)
\\
&= \frac{c\varkappa^{2}}{8}\lt[\sV t_{\bxi}\re-\qV(t\re_{\bxi})(t-t\re_{\bxi})-2\int_{0}^{t\re_{\bxi}}\qV(t')\drm{t'} \rt]+
O(\varkappa^4)
\end{split}
\end{align}
Inserting \req{14} in \req{4b} and \req{4l}, and \req{15} in \req{4e} and \req{4k}, and
\req{16} in \req{4h}, and \req{17} in \req{4f}, and finally
\req{18} in \req{4c} and \req{4i}, the constituent $\bsym{\pi}_{\bxi,(4)}^{[k],i}$ terms become
\begin{align} %eq:19
\label{eq:pi014}
&\bsym{\pi}_{\bxi,(4)}^{[0],1}(t,\sV)
=-\frac{\varkappa^{4}}{4}\lt[\frac{\{(\nh-\bsym{\beta})\cdot \nh\} \bsym{\beta}-\{(\nh-\bsym{\beta})\cdot \bsym{\beta} \}\nh }{(1-\bsym{\beta}\cdot \nh)^{2}}\rt]\re
\\
\label{eq:pi124}
&\bsym{\pi}_{\bxi,(4)}^{[1],2}(t,\sV)=
\frac{\varkappa^{4}}{8c^{2}}
\lt[\frac{
\{({\color{red}\aV}\cdot \sV)-(\nh\cdot {\color{red}\aV})(\nh \cdot\sV)\}\qV-\{({\color{red}\aV}\cdot \qV)-(\nh\cdot {\color{red}\aV})(\nh \cdot\qV)\}\sV}
{(1-\bsym{\beta}\cdot \nh )^{3}}\rt]\re
\\
\notag
&\bsym{\pi}_{\bxi,(4)}^{[1],3}(t,\sV)=
\frac{\varkappa^{4}}{8c}
\Biggl[\frac{\{t\re_{\bxi}({\color{red}\aV}\cdot \sV)-(t-t\re_{\bxi})({\color{red}\aV}\cdot \qV)\}\nh-\{t\re_{\bxi}(\nh \cdot\sV)-(t-t\re_{\bxi})(\nh \cdot\qV)\}{\color{red}\aV}}
{(1-\bsym{\beta}\cdot \nh )^{3}}\Biggr]\re
\\ \label{eq:pi134}
&\qquad\qquad
-\frac{\varkappa^{4}}{4c}\lt[\frac{\displaystyle{\bigg({\color{red}\aV}\cdot\int_{0}^{t\re_{\bxi}} \qV(t')\drm{t'}\bigg)\nh
-\bigg(\nh\cdot\int_{0}^{t\re_{\bxi}} \qV(t')\drm{t'}\bigg){\color{red}\aV}}}
{(1-\bsym{\beta}\cdot \nh )^{3}}\rt]\re
\\
\label{eq:pi154}
&\bsym{\pi}_{\bxi,(4)}^{[1],5}(t,\sV)=
-\frac{\varkappa^{4}}{8c^{2}}
(ct-|\sV|)
\lt[\frac{({\color{red}\aV}\cdot \sV)\nh - (\nh\cdot \sV){\color{red}\aV}}{(1-\bsym{\beta}\cdot \nh )^{3}}\rt]\re
\\
\label{eq:pi164}
&\bsym{\pi}_{\bxi,(4)}^{[1],6}(t,\sV)
=
\frac{\varkappa^{4}}{2}
\lt[\frac{ t\re_{\bxi}\vV +\qV}{(1-\bsym{\beta}\cdot \nh)}\rt]\re
\\
\label{eq:pi184}
&\bsym{\pi}_{\bxi,(4)}^{[1],8}(t,\sV)=
-
\frac{\varkappa^{4}}{2} (ct-|\sV|)\lt[\frac{
\bsym{\beta}}{1-\bsym{\beta}\cdot \nh}\rt]\re,
\end{align}
and for the $\bsym{\pi}_{\bxi,(4)}^{[2]}$ terms we find
\begin{align}%eq:19
\label{eq:pi224}
&\hspace{-1truecm}
\bsym{\pi}_{\bxi,(4)}^{[2],2}(t,\sV)
={+}\frac{\varkappa^{4}}{8}\lt[\frac{c(1-\beta^{2})}{(1-\bsym{\beta}\cdot \nh )^{3}}\rt]\re
\Bigg[\{(\bsym{\beta}\cdot \sV)t\re_{\bxi}-(t-t\re_{\bxi})(\qV\cdot \bsym{\beta})\}\nh 
\\ \notag
 & \hspace{5.2truecm} -\{(\sV\cdot \nh)t\re_{\bxi}-(t-t\re_{\bxi})(\qV\cdot \nh)\}\bsym{\beta}
\\ \notag
& \hspace{5.2truecm} -2\lt(\bsym{\beta}\cdot\int_{0}^{t\re_{\bxi}}\! \qV(t')\drm{t}^\prime\rt)\nh
 +2\lt(\nh\cdot\int_{0}^{t\re_{\bxi}}\!\qV(t')\drm{t}^\prime \rt)\bsym{\beta}\Bigg]\re
\\
\label{eq:pi244}
&\hspace{-1truecm}
\bsym{\pi}_{\bxi,(4)}^{[2],4}(t,\sV)=
{-}
\frac{\varkappa^{4}}{8}(ct-|\sV|)\lt[\frac{ (1-\beta^{2})\lt[(\sV\cdot \bsym{\beta})\nh-(\nh \cdot \sV)\bsym{\beta}\rt]}
{(1-\bsym{\beta}\cdot \nh )^{3}}\rt]\re
\\
\label{eq:pi254}
&\hspace{-1truecm}
\bsym{\pi}_{\bxi,(4)}^{[2],5}(t,\sV)
=-\varkappa^{4}\frac{1}{8}\lt[\frac{(1-\beta^{2})
\lt[\{(  \nh-\bsym{\beta})\cdot \sV\}\qV-\{( \nh-\bsym{\beta})\cdot \qV\}\sV\rt]}{
(1-\bsym{\beta}\cdot \nh )^{3}}\rt]\re
\end{align}

 To evaluate the integral \req{Zdef} we need the $\bsym{\pi}_{\bxi,(4)}^{[k],j}(t,\sV)$
with $\sV=\qV(t^{r})+c(t-t^{r})\nh$, which lets us solve for the retarded time 
$t\re_{\bxi}=t\re_{\bxi}(t,\sV)$ defined by $c(t-t\re_{\bxi})=|\sV-\qV(t\re_{\bxi})|$. 
Hence,
\begin{equation}
\label{eq:20}
c(t-t\re_{\bxi})=|\qV(t^{r})-\qV(t\re_{\bxi})+c(t-t^{r})\nh|,
\end{equation} 
which is evidently solved by $t\re_{\bxi}=t^{r}$.
 With this in mind we do away with the $t^{\mbox{\tiny{ret}}}_{\bxi}$ notation. 
 When putting the terms together and evaluating $\bsym{\pi}_{\bxi}^{[k]}$ we find that terms containing 
factors $|\sV|$ cancel happily. 
 Other terms are collected under common factors and we have
\begin{align}
\label{eq:21}
\bsym{\pi}_{\bxi{,(4)}}^{[0]}=&-\frac{\varkappa^{4}}{4}\lt[\frac{\{(\nh-\bsym{\beta}(t^{r}))\cdot \nh\} 
\bsym{\beta}(t^{r})-\{(\nh-\bsym{\beta}(t^{r}))\cdot \bsym{\beta}(t^{r}) \}\nh }{(1-\bsym{\beta}(t^{r})\cdot \nh)^{2}}\rt],
\\
\notag
\bsym{\pi}_{\bxi{,(4)}}^{[1]}
=&\frac{\varkappa^{4}}{8c^{2}}\frac{1}{(1-\bsym{\beta}(t^{r})\cdot \nh )^{3}}
\Bigg[-c(t-t^{r})\{{\color{red}\aV}(t^{r})\cdot (\qV(t^{r}) +c(t-t^{r})\nh)+{\color{red}\aV}(t^{r})\cdot \qV(t^{r})\}\nh
\\ \notag
&\qquad \qquad \qquad \qquad 
+c(t-t^{r})\{\nh \cdot(\qV(t^{r})+c(t-t^{r})\nh)+\nh \cdot\qV(t^{r})\}{\color{red}\aV}(t^{r})
\\ \notag
&\qquad \qquad \qquad \qquad 
-2\lt({\color{red}\aV}(t^{r})\cdot\int_{0}^{t^{r}} \qV(t')\drm{t}^\prime\rt)\nh+2\lt(\nh\cdot\int_{0}^{t^{r}}
\qV(t')\drm{t}^\prime \rt){\color{red}\aV}(t^{r})
\\ \notag
&\qquad 
+\{{\color{red}\aV}(t^{r})\cdot (\qV(t^{r})+c(t-t^{r})\nh)-(\nh\cdot {\color{red}\aV}(t^{r}))(\nh \cdot(\qV(t^{r})+c(t-t^{r})\nh))\}\qV(t^{r})
\\ \notag
&\qquad \qquad \qquad 
-\{({\color{red}\aV}(t^{r})\cdot \qV(t^{r}))-(\nh\cdot {\color{red}\aV}(t^{r}))(\nh \cdot\qV(t^{r}))\}(\qV(t^{r})+c(t-t^{r})\nh)\Bigg]
\\ \label{eq:21b}
&+\frac{\varkappa^{4}}{2}\frac{\qV(t^{r})-\vV(t^{r})(t-t^{r})}{(1-\bsym{\beta}(t^{r})\cdot \nh)},
\end{align}
\begin{align}
 \notag
\bsym{\pi}_{\bxi{,(4)}}^{[2]}
=&{+}\frac{\varkappa^{4}}{8}\frac{1-\beta^{2}(t^{r})}{(1-\bsym{\beta}(t^{r})\cdot \nh )^{3}}
\Bigg[-c(t-t^{r})\{\bsym{\beta}(t^{r})\cdot (\qV(t^{r})+c(t-t^{r})\nh)+\bsym{\beta}(t^{r})\cdot \qV(t^{r})\}\nh
\\ \notag
&\qquad \qquad \qquad \qquad \qquad 
+c(t-t^{r})\{\nh \cdot (\qV(t^{r})+c(t-t^{r})\nh)+\nh \cdot\qV(t^{r})\}\bsym{\beta}(t^{r})
\\  \notag
&\hspace{-1truecm}
{-}\{(\nh-\bsym{\beta}(t^{r}))\cdot (\qV(t^{r})+c(t-t^{r})\nh)\}\qV(t^{r})
{+}\{(\nh-\bsym{\beta}(t^{r}))\cdot \qV(t^{r})\}(\qV(t^{r})+c(t-t^{r})\nh)
\\ \label{eq:22}
&\qquad \qquad \qquad \qquad \quad 
+2\lt(\bsym{\beta}(t^{r})\cdot\!\int_{0}^{t^{r}}\!\! \qV(t')\drm{t}^\prime\rt)\nh-2\lt(\nh\cdot\!\int_{0}^{t^{r}}\!\!
\qV(t')\drm{t}^\prime \rt)\bsym{\beta}(t^{r})\Bigg],
\end{align}
where we have inserted $\sV=\qV(t^{r})+c(t-t^{r})\nh$. 
Collecting like terms and canceling where appropriate, and using that in $\sV=\qV(t^{r})+c(t-t^{r})\nh$ the unit 
vector $\nh(\qV,\sV)$ is simply 
$\nV =\left(\sin\vartheta \cos\varphi,\;\sin\vartheta \sin\varphi ,\; \cos\vartheta \right)$, {so that
$\bsym{\beta}(t^r)\cdot\nV=\beta(t^r)\cos\vartheta$}, {and setting 
${\color{red}\aV}\cdot \nh= {\color{red}a}\cos\vartheta $, $\qV\cdot \nh= q\cos\vartheta $ and 
$\qV\cdot {\color{red}\aV}={\color{red}a}q$,} we get
\begin{align}
\label{eq:22b}
\hspace{-1truecm}
\bsym{\pi}_{\bxi{,(4)}}^{[0]}
=&
-\frac{\varkappa^{4}}{4}\lt[\frac{(1-\beta\cos \vartheta) \bsym{\beta}-(\beta \cos \vartheta -\beta^{2})\nh }{(1-\beta\cos \vartheta)^{2}}\rt]\re
\\ \notag
\hspace{-1truecm}
\bsym{\pi}_{\bxi{,(4)}}^{[1]}
=& 
\frac{\varkappa^{4}}{8c^{2}(1-\beta(t^{r})\cos \vartheta )^{3}}\Bigg[-c(t-t^{r})\{3
q(t^{r})+c(t-t^{r}) \cos \vartheta- \cos^{2} \vartheta(t^{r}) q(t^{r})\}a(t^{r})\nh
\\ \notag
\hspace{-1truecm}&\qquad \qquad \qquad \qquad \qquad 
+c(t-t^{r})\{2 \cos \vartheta q(t^{r})+c(t-t^{r})\}{\color{red}\aV}(t^{r}) 
\\ \notag
\hspace{-1truecm} &
\qquad \qquad \qquad \qquad \qquad  -2\lt(a(t^{r})\int_{0}^{t^{r}}\!
q(t')\drm{t}^\prime \rt)\nh+2\lt(\cos \vartheta \int_{0}^{t^{r}}\!q(t')\drm{t}^\prime \rt){\color{red}\aV}(t^{r})\Bigg]
\\ \label{eq:22bb}
\hspace{-1truecm}
&+\frac{\varkappa^{4}}{2}\frac{{\qV(t^{r})-\vV(t^{r})(t-t^{r})}}{1-\beta(t^{r})\cos \vartheta}
\\ \notag
\hspace{-1truecm}
\bsym{\pi}_{\bxi{,(4)}}^{[2]}
=& {+}
\frac{\varkappa^{4}}{8}
\frac{1-\beta^{2}(t^{r})}{(1-\beta(t^{r})\cos \vartheta )^{3}}\Bigg[-\{ {3}\beta(t^{r})q(t^{r})
{-}(q(t^r){-}v(t^r)(t-t^r))\cos \vartheta\}c(t-t^{r})\nh
\\ \notag
\hspace{-1truecm}
&\qquad +\{c(t-t^{r})+2 q(t^{r})\cos \vartheta\}c(t-t^{r})\bsym{\beta}(t^{r})
{-}\{(1-\beta(t^{r})\cos \vartheta) \}c(t-t^{r})\qV(t^r)
\\ \label{eq:22bbb}
\hspace{-1truecm}
& \qquad \qquad \qquad \qquad  
+2\lt(\beta(t^{r})\int_{0}^{t^{r}}\!q(t')\drm{t}^\prime \rt)\nh-2\lt(\cos\vartheta\int_{0}^{t^{r}}\!q(t')\drm{t}^\prime\rt)\bsym{\beta}(t^{r})\Bigg]
\end{align}
%%%%%%%%%%%%%%%%%%%%%%%%%%%%%%%%%%%%%%%%%
\subsection{Angular Integrals}
Armed with the $\bsym{\pi}_{\bxi{,(4)}}^{[k]}$ kernels, we are in a position to evaluate the angular integrals that yield the 
4th order contributions of the $\bld{Z}_{\bxi}^{[k]}$ terms.
{Having displayed all the dependencies on $t^r$ explicitly above, we from now on simplify notation and 
understand that, unless stated explicitly otherwise, we have $\bsym{\beta} \equiv \bsym{\beta}(t^{r})$ and 
$\beta\equiv \beta(t^r)$ and $\qV\equiv\qV(t^r)$ as well as ${\color{red}\aV}\equiv{\color{red}\aV}(t^r)$.}

 When $k=0$, we have 
\begin{align}
\begin{split}
\label{eq:23}
\bld{Z}_{\bxi}^{[0],(4)}(t,t^{r}) = 
-\frac{\varkappa^{4}}{4}\int^{2\pi}_{0}\int^{\pi}_{0} 
 \lt[\frac{(1-\beta \cos\vartheta)\bsym{\beta}-\beta(\cos\vartheta-\beta)\nh}{1-\beta \cos \vartheta}\rt]\sin\vartheta \drm\vartheta\drm\varphi.
\end{split}
\end{align}
Here, the superscript ${}^{(4)}$ denotes that this is the $4$th order contribution to $\bld{Z}^{[0]}_{\bxi}$. 
{Note that while $\bld{Z}_{\bxi}^{[0]}(t,t^{r})$ does in general depend on both $t$ and $t^r$, 
$\bld{Z}_{\bxi}^{[0],(4)}(t,t^{r})$ is independent of $t$.}
When evaluating the integral over $\varphi$, 
we observe that the only dependence is on the unit vector $\nh$ in the directions perpendicular to the motion. 
The integrations of these components will vanish and for the parallel component we simply obtain a factor of $2\pi$. 
Denoting the parallel component  by $\parallel$, we find
\begin{align}
\begin{split}
\label{eq:24}
Z^{[0],(4)}_{{\bxi,}\parallel}(t,t^{r})=
-\varkappa^{4}\frac{\pi}{2}\int^{\pi}_{0} 
 \lt[\beta-\frac{\beta(\cos\vartheta-\beta)\cos \vartheta}{1-\beta \cos \vartheta}\rt]\sin\vartheta \drm\vartheta.
\end{split}
\end{align}
The integral can be carried out explicitly, yielding
\begin{align}
\label{eq:25}
Z^{[0],(4)}_{{\bxi,}\parallel}(t,t^{r})
= -\varkappa^{4}\frac{\pi}{2}\lt[\frac{2}{\beta}+\lt(1-\frac{1}{\beta^{2}}\rt)\ln\lt(\frac{1+\beta}{1-\beta}\rt)\rt].
\end{align}

 The next term is for $k=1$. 
We have
\begin{align}
\notag
\hspace{-.5truecm}
\bld{Z}^{[1],(4)}_{{\bxi}}(t,t^{r})=&\frac{\varkappa^{4}}{8c^{2}}\int^{2\pi}_{0}
\int^{\pi}_{0} 
\frac{\sin \vartheta}{(1-\beta\cos\vartheta)^{2}}\Bigg[-\lt(3q+c(t-t^{r}) \cos\vartheta-q\cos^{2}\vartheta\rt){\color{red}a}c(t-t^{r})\nh
\\ \notag
\hspace{-.5truecm}
&\qquad \qquad \qquad \qquad 
+\lt(2q\cos\vartheta+c(t-t^{r})\rt)c(t-t^{r}){\color{red}\aV}
\\ \notag
\hspace{-.5truecm}
&\qquad \qquad -2\lt({\color{red}\aV}\cdot\int_{0}^{t^r} \qV(t')\drm{t}^\prime\rt)\nh+
2\lt(\nh\cdot\int_{0}^{t^r} \qV(t')\drm{t}^\prime\rt){\color{red}\aV}\Bigg]\drm\vartheta\drm\varphi
\\
\hspace{-.5truecm}
& +\frac{\varkappa^{4}}{2}\int^{2\pi}_{0}
\int^{\pi}_{0} \sin \vartheta \lt(\qV-(t-t^{r})\vV\rt)\drm\vartheta \drm\varphi,
\label{eq:26}
\end{align}
where ${\color{red}a}$ and $q$ are tacitly functions of $t^r$.
 Like before we find that all components not aligned with the motion vanish, while the $\varphi$ integrals 
contribute a factor of $2\pi$ to the aligned terms.
 We can easily evaluate the last term, such that the component parallel to the motion is
\begin{align}
\begin{split}
\label{eq:27}
\hspace{-1truecm}
Z^{[1],(4)}_{{\bxi,}||}(t,t^{r})=\varkappa^{4}\frac{\pi}{4c^{2}}
\int^{\pi}_{0} &
\frac{\sin \vartheta}{(1-\beta \cos\vartheta)^{2}}\times \\ 
& \times \Bigg[-\lt(c(t-t^{r}) \cos\vartheta +q(3-\cos^{2}\vartheta)\rt){\color{red}a}c(t-t^{r})\cos \vartheta
\\
&\qquad  +\lt(2q\cos\vartheta+c(t-t^{r})\rt)c(t-t^{r}){\color{red}a}
\\
&\qquad 
-2\lt(a\int_{0}^{t^r}\! q(t')\drm{t}^\prime\rt)\cos \vartheta+2\lt(\cos\vartheta\int_{0}^{t^r}\!q(t')\drm{t}^\prime\rt){\color{red}a}\Bigg]\drm\vartheta
\\
&\hspace{-1.5truecm} + \varkappa^{4}2 \pi  \lt(q-v(t-t^{r})\rt).
\end{split}
\end{align}
Notably the terms on the third line of the $[\cdots]$ bracket cancel out. 
Collecting like terms
\begin{align}
\begin{split}
\label{eq:28}
Z^{[1],(4)}_{{\bxi,}||}(t,t^{r})=&\varkappa^{4}\frac{\pi}{4c^{2}}
\int^{\pi}_{0}\frac{\sin \vartheta}{(1-\beta \cos\vartheta)^{2}}\Bigg[{\color{red}a}qc(t-t^{r})(\cos^{3}\vartheta-\cos \vartheta)
\\
& \hspace{4.5truecm} + {\color{red}a}c^2(t-t^{r})^{2}\lt(1-\cos^{2}\vartheta\rt)\Bigg] \drm\vartheta \\ 
& +\varkappa^{4}2 \pi  \lt(q-v(t-t^{r})\rt).
\end{split}
\end{align}
 When evaluating over $\vartheta$ we can break up the integration into a number of terms that have similar integrals. 
 We evaluate 
\begin{align}
\begin{split}
\label{eq:29}
&\int_{0}^{\pi}\frac{\sin \vartheta}{(1-\beta \cos\vartheta)^{2}}\drm\vartheta
=\frac{2}{(1-\beta^{2})},
\\
&\int_{0}^{\pi} \frac{\sin \vartheta\cos \vartheta}{(1-\beta \cos\vartheta)^{2}}\drm\vartheta
=\frac{2}{\beta(1-\beta^{2})}-\frac{1}{\beta^{2}}\ln\lt(\frac{1+\beta}{1-\beta}\rt),
\\
&\int_{0}^{\pi}\frac{\sin \vartheta\cos^{2} \vartheta}{(1-\beta \cos\vartheta)^{2}}\drm\vartheta
=\frac{2}{\beta^{2}(1-\beta^{2})}-\frac{2}{\beta^{3}}\ln\lt(\frac{1+\beta}{1-\beta}\rt)+\frac{2}{\beta^{2}},
\\
&\int_{0}^{\pi}\frac{\sin \vartheta\cos^{3} \vartheta}{(1-\beta \cos\vartheta)^{2}}\drm\vartheta
=\frac{2}{\beta^{3}(1-\beta^{2})}-\frac{3}{\beta^{4}}\ln\lt(\frac{1+\beta}{1-\beta}\rt)+\frac{4}{\beta^{3}}.
\end{split}
\end{align}
Hence
\begin{align}
\begin{split}
\label{eq:31}
Z^{[1],(4)}_{{\bxi,}||}(t,t^{r})
=\varkappa^{4}\frac{\pi}{4c^{2}}\Bigg[&{\color{red}a}qc(t-t^{r})
\Bigg(\frac{6}{\beta^{3}}+\frac{1-\frac{3}{\beta^{2}}}{\beta^{2}}
\ln\Big(\frac{1+\beta}{1-\beta}\Big)\!\!\Bigg)\!
\\
&+{\color{red}a}c^2(t-t^{r})^2\lt(-\frac{4}{\beta^{2}}+\frac{2}{\beta^{3}}\ln\lt(\frac{1+\beta}{1-\beta}\rt)\!\! \rt)\!\!\Bigg]
\\
&\hspace{-1.5truecm} +\varkappa^{4} 2 \pi  \lt(q-v(t-t^{r})\rt).
\end{split}
\end{align}

The next term is 
\begin{align}
\begin{split}
\label{eq:32}
\bld{Z}^{[2],(4)}_{{\bxi}}(t,t^{r})=
&\varkappa^{4}\frac18
\Big(1-\beta^{2})\Big)\int^{2\pi}_{0}\!\int^{\pi}_{0} 
\frac{\sin \vartheta}{(1-\beta \cos\vartheta)^{2}}\times\\ 
&\times\Bigg[{-}\{{3}\bsym{\beta}\cdot \qV
{-}
(\qV{-}(t-t^r)\vV)\cdot \nh\}c(t-t^{r})\nh
\\
&\qquad 
{+}\{c(t-t^{r})+2\nh \cdot \qV\}c(t-t^{r})\bsym{\beta}-\{(\nh-\bsym{\beta})\cdot \nh \}c(t-t^{r})\qV
\\
&\qquad {+}2\lt(\bsym{\beta}\cdot\int_{0}^{t^{r}}\!
\qV(t')\drm{t}^\prime \rt)\nh
{-}2\lt(\nh\cdot\int_{0}^{t^{r}}\! \qV(t')\drm{t}^\prime\rt)\bsym{\beta}\Bigg]\drm\vartheta \drm\varphi.
\end{split}
\end{align}
Like before, we integrate over $\varphi$, eliminating all components that are not parallel to the motion. 
We find that {after $\varphi$ integration}
 the two terms containing integrals over $t'$ cancel,
as well as terms sharing the coefficient {$\beta q \cos\vartheta$}, thus
\begin{align}
\begin{split}
\label{eq:33}
Z^{[2],(4)}_{{\bxi,}\parallel}(t,t^{r})=&\varkappa^{4}\frac{\pi}{4}(1-\beta^{2})\int^{\pi}_{0} 
\frac{\lt[q c(t-t^{r}) {-}\beta c^{2}(t-t^{r})^{2}\rt]\sin \vartheta}{(1-\beta \cos\vartheta)^2}
\Big[\cos^{2} \vartheta-1\Big] \drm\vartheta.
\end{split}
\end{align}
Integration over $\vartheta$, {using (\ref{eq:29}),}  simply yields
\begin{align}
\label{eq:34}
Z^{[2],(4)}_{{\bxi,}\parallel}(t,t^{r})=
\varkappa^{4}\frac{\pi}{4} \Big(1-\beta^{2}\Big)\!
\Big[qc(t-t^{r}){-}\beta c^{2}(t-t^{r})^{2}
 \Big]\! \Bigg[\frac{4}{\beta^{2}}-\frac{2}{\beta^{3}}\ln 
\lt(\frac{1+\beta}{1-\beta}\rt)\!\!\Bigg].
\end{align}

According to \req{selfFexpl}, the self-force also depends on the time derivative of these terms.
 We readily find
\begin{align}
\label{eq:35a}
\pt{t} \bld{Z}^{[0],(4)}_{\bxi}(t,t^{r})&=0,
\\
\label{eq:35b}
\pt{t} \bld{Z}^{[1],(4)}_{\bxi}(t,t^{r})&=\varkappa^{4}\frac{\pi}{8c^{2}}
\Bigg[{\color{red}a}qc\Bigg(\frac{6}{\beta^{3}}+\frac{1-\frac{3}{\beta^{2}}}{\beta^{2}}\ln\lt(\frac{1+\beta}{1-\beta}\rt)\!\!\!\Bigg)
\!\\ \notag
&\qquad\qquad +2{\color{red}a}c^{2}(t-t^{r})\!\lt(-\frac{4}{\beta^{2}}+\frac{2}{\beta^{3}}\ln\lt(\frac{1+\beta}{1-\beta}\rt)\!\! \rt)\!\!
\Bigg]\be - \varkappa^{4} 2 \pi\vV,
\\
\label{eq:35c}
\pt{t} \bld{Z}^{[2],(4)}_{\bxi}(t,t^{r})&=\varkappa^{4}\frac{\pi}{4}(1-\beta^{2})
\lt[qc{-}2\beta c^{2}(t-t^{r}) \rt]
\lt(\!\frac{4}{\beta^{2}}-\frac{2}{\beta^{3}}\ln \lt(\frac{1+\beta}{1-\beta}\rt)\!\!\rt)\!\!\be;
\end{align}
recall that $\be$ is the unit vector pointing along $\EV^{\mbox{\tiny{hom}}}$, thus $\be=\qV/|\qV|$ too. 

According to \req{selfFexpl}, the self-force also depends on $Z^{[2],4}_{{\bxi,}||}(t,t)=\lim_{t^r\to t} Z^{[2],4}_{{\bxi,}||}(t,t^r)$. 
From \req{34} we readily find that $ Z^{[2],4}_{{\bxi,}||}(t,t) \equiv 0 $, making no contribution.

\subsection{The final formula for the self-force at $O(\varkappa^4)$}
 We now have all the ingredients needed to evaluate the self-force term \req{selfFexpl} at 4th order in powers of $\varkappa$.
 We denote the 4th order term as $\bld{F}_{0}^{(4)}(t)=F_{0}^{(4)}(t)\be$.
 Since all contributing terms are $\propto\varkappa^4 {e^2}$, 
we can factor out an overall coefficient $\varkappa^4 {e^2}$. 
 After also canceling $\pi$s wherever possible, {$F_{0}^{(4)}(t)$} is given by
\begin{align}
\begin{split}
\label{eq:36}
F_{0}^{(4)}(t)=\frac{1}{4}\varkappa^{4}e^{2}\Bigg[
& c^{2}
\int_{0}^{t}(t-t^{r})\lt[\frac{2}{\beta}+\lt(1-\frac{1}{\beta^{2}}\rt)\ln\lt(\frac{1+\beta}{1-\beta}\rt)\rt]\drm{t}^r
\\
&-\frac{1}{4} \int_{0}^{t}
\Bigg[(t-t^{r})\Bigg(\frac{6}{\beta^{3}}+\frac{1-\frac{3}{\beta^{2}}}{\beta^{2}}\ln\lt(\frac{1+\beta}{1-\beta}\rt)\Bigg)q{\color{red}a}
\\
&\hspace{3truecm} 
+c(t-t^{r})^2\lt(-\frac{4}{\beta^{2}}+\frac{2}{\beta^{3}}\ln\lt(\frac{1+\beta}{1-\beta}\rt) \rt){\color{red}a}\Bigg]\drm{t}^r
\\
& -2  c  \int_{0}^{t} \lt(q-(t-t^{r})v\rt) \drm{t}^r
\\
&-\frac{1}{4}\int_{0}^{t}
{\Bigg[(t-t^{r})}
{\color{red}a}
q\Bigg(\frac{6}{\beta^{3}}+\frac{1-\frac{3}{\beta^{2}}}{\beta^{2}}\ln\lt(\frac{1+\beta}{1-\beta}\rt)\Bigg)
\\
&\hspace{3truecm}
+ 2c(t-t^{r})^{2}\lt(-\frac{4}{\beta^{2}}+\frac{2}{\beta^{3}}\ln\lt(\frac{1+\beta}{1-\beta}\rt) \rt){\color{red}a}\Bigg]\drm{t}^r
\\
&+2  c \int_{0}^{t} (t-t^{r})v\drm{t}^r
\\
&{+} c^{2}\int_{0}^{t}
(t-t^{r})(1-\beta^{2})\Bigg[\frac{2}{\beta}-\frac{1}{\beta^{2}}\ln \lt(\frac{1+\beta}{1-\beta}\rt)\Bigg]\drm{t}^r
\\
&-\frac{1}{4}\int_{0}^{t}
qc(1-\beta^{2})\Bigg[\frac{4}{\beta^{2}}-\frac{2}{\beta^{3}}\ln \lt(\frac{1+\beta}{1-\beta}\rt)\Bigg]\Bigg]\drm{t}^r.
\end{split}
\end{align}
There are many terms that can be collected such as the 
second and fifth lines {(that merge into the third line of the next formula)}; the 
third and sixth lines can also be collected {(which merge into the fourth line of the next formula)};
lines one and eight may also be collected together {(featuring in the first two lines of the next formula)}. 
Line four simply evaluates to zero via integration by parts. 

 This yields
\begin{align}
\begin{split}
\label{eq:37}
F_{0}^{(4)}(t)=\frac{1}{4}\varkappa^{4}e^{2}\Bigg[& {-}c^{2}\int_{0}^{t}(t-t^{r})2\beta \drm{t}^r
\\
& { +2 c^{2}
\int_{0}^{t}(t-t^{r})\lt[\frac{2}{\beta}+\lt(1-\frac{1}{\beta^{2}}\rt)\ln\lt(\frac{1+\beta}{1-\beta}\rt)\rt]\drm{t}^r }
\\
&-\frac{1}{4}\int_{0}^{t}
\Bigg[2(t-t^{r})\Bigg(\frac{6}{\beta^{3}}+\frac{1-\frac{3}{\beta^{2}}}{\beta^{2}}\ln\lt(\frac{1+\beta}{1-\beta}\rt)\Bigg)q
\\
&\hspace{2truecm}
+3c(t-t^{r})^{2}\lt(-\frac{4}{\beta^{2}}+\frac{2}{\beta^{3}}\ln\lt(\frac{1+\beta}{1-\beta}\rt) \rt)\Bigg]{\color{red}a} \drm{t}^r
\\
&+2  c \int_{0}^{t} (t-t^{r}) v\drm{t}^r
\\
&-\frac{1}{4}\int_{0}^{t}
qc(1-\beta^{2})\lt(\frac{4}{\beta^{2}}-\frac{2}{\beta^{3}}\ln \lt(\frac{1+\beta}{1-\beta}\rt)\!\!\rt)\drm{t}^r\Bigg].
\end{split}
\end{align}
{The first and fifth lines cancel.
 On the fourth line we can perform integration by parts; thus}
\begin{align}
\begin{split}
\label{eq:38}
c\int_{0}^{t}
(t-t^{r})^{2}&\lt(-\frac{4}{\beta^{2}}+\frac{2}{\beta^{3}}\ln\lt(\frac{1+\beta}{1-\beta}\rt) \rt){\color{red}a} \drm{t}^r
=\\
& c^{2}(t-t^{r})^{2}\int_{0}^{t^{r}}
\frac{d\beta}{d\tau}\lt(-\frac{4}{\beta^{2}}+\frac{2}{\beta^{3}}\ln\lt(\frac{1+\beta}{1-\beta}\rt)\rt)\drm\tau\Bigg|_{t^r=0}^{t^r=t}
\\
&+2c^{2}\int_{0}^{t} 
(t-t^{r})\int_{0}^{\beta}\lt(-\frac{4}{x^{2}}+\frac{2}{x^{3}}\ln\lt(\frac{1+x}{1-x}\rt)\rt)
\drm{x}\drm{t}^r.
\end{split}
\end{align}
The first line at r.h.s.\req{38} vanishes whereas the second can be evaluated, in total yielding
\begin{align}
\begin{split}
\label{eq:39}
c\int_{0}^{t}(t-t^{r})^{2}&\lt(-\frac{4}{\beta^{2}}+\frac{2}{\beta^{3}}\ln\lt(\frac{1+\beta}{1-\beta}\rt) \rt){\color{red}a}\drm{t}^r\\
& \qquad 
= 2c^{2}\int_{0}^{t} (t-t^{r})\lt(\frac{2}{\beta}+\lt(1-\frac{1}{\beta^{2}}\rt)\ln\lt(\frac{1+\beta}{1-\beta}\rt)\!\rt)\drm{t}^r,
\end{split}
\end{align}
{which is the same expression as the one in the second line of (\ref{eq:37}); 
of course, (\ref{eq:39}) still needs to be multiplied by $-3/4$ before being added to the 
second line of (\ref{eq:37}).
 We arrive at}
\begin{align}
\begin{split}
\label{eq:40}
F_{0}^{(4)}(t)= \frac{1}{4}\varkappa^{4}e^{2}\Bigg[
& -\frac{1}{4}\int_{0}^{t}
\Bigg[2(t-t^{r})\Bigg(\frac{6}{\beta^{3}}+\frac{1-\frac{3}{\beta^{2}}}{\beta^{2}}\ln\lt(\frac{1+\beta}{1-\beta}\rt)\Bigg)\Bigg]q{\color{red}a}
\drm{t}^r
\\
&
{
+\frac{1}{2}}c^{2}\int_{0}^{t}
(t-t^{r})\lt(\frac{2}{\beta}+\lt(1-\frac{1}{\beta^{2}}\rt)\ln\lt(\frac{1+\beta}{1-\beta}\rt)\rt) \drm{t}^r
\\
& -\frac{1}{4}\int_{0}^{t}
(1-\beta^{2})qc\lt(\frac{4}{\beta^{2}}-\frac{2}{\beta^{3}}\ln \lt(\frac{1+\beta}{1-\beta}\rt)\!\rt)\drm{t}^r\Bigg]
\end{split}
\end{align}
 We can also integrate by parts the remaining term containing ${\color{red}a}$, {thus}
\begin{align}
\begin{split}
\label{eq:41}
&\int_{0}^{t}
(t-t^{r})\Bigg(\frac{6}{\beta^{3}}+\frac{1-\frac{3}{\beta^{2}}}{\beta^{2}}\ln \lt(\frac{1+\beta}{1-\beta}\rt)\!\Bigg)q{\color{red}a}\drm{t}^r
\\
&=qc(t-t^{r})
\int_{0}^{t^{r}}
\frac{d\beta}{d\tau}\Bigg(\frac{6}{\beta^{3}}+\frac{1-\frac{3}{\beta^{2}}}{\beta^{2}}\ln\lt(\frac{1+\beta}{1-\beta}\rt)\!\Bigg)
\!(\tau) \drm\tau \Bigg|_{t^r=0}^{t^r=t}
\\
&-c\int_{0}^{t}
\frac{d}{\drm{t}^r}\lt(q(t-t^{r})\rt)\int_{0}^{\beta}\lt(\frac{6}{x^{3}}+\frac{\lt(1-\frac{3}{x^{2}}\rt)}{x^{2}}\ln  
\lt(\frac{1+x}{1-x}\rt)\rt)\drm{x}\drm{t}^r.
\end{split}
\end{align}
The boundary term vanishes while the inner integral over $x$ can be evaluated, and so
\begin{align}
\notag
&\int_{0}^{t}
(t-t^{r})\Bigg(\frac{6}{\beta^{3}}+\frac{1-\frac{3}{\beta^{2}}}{\beta^{2}}\ln  \lt(\frac{1+\beta}{1-\beta}\rt)\!\Bigg)q{\color{red}a}\drm{t}^r
\\ \label{eq:42a}
&=-c\int_{0}^{t}\lt(c (t-t^{r})\beta-q\rt)\lt(\frac{4}{3}-\frac{1}{\beta}\lt[\frac{2}{\beta}+\lt(1-\frac{1}{\beta^{2}}\rt)\ln\lt(\frac{1+\beta}{1-\beta}\rt)\rt]\rt)\drm{t}^r
\\ \label{eq:42b}
&=-\frac{4c}{3}\int_{0}^{t} \lt(c (t-t^{r})\beta-q\rt)\drm{t}^r
+c^{2}\int_{0}^{t}
(t-t^{r})\lt[\frac{2}{\beta}+\lt(1-\frac{1}{\beta^{2}}\rt)\ln\lt(\frac{1+\beta}{1-\beta}\rt)\rt]\drm{t}^r
\\ \notag
&\qquad -c\int_{0}^{t}
\frac{q}{\beta}\lt[\frac{2}{\beta}+\lt(1-\frac{1}{\beta^{2}}\rt)\ln\lt(\frac{1+\beta}{1-\beta}\rt)\rt]\drm{t}^r
\end{align}
and the integral over $c\beta (t-t^{r})-q$ at r.h.s.\req{42b} vanishes.
 The self-force at $O(\varkappa^4)$ becomes
\begin{align}
\begin{split}
\label{eq:43}
F_{0}^{(4)}(t)=\frac{1}{4}\varkappa^{4}e^{2}\Bigg[&-\frac{1}{2}c^{2}\int_{0}^{t}
(t-t^{r})\lt[\frac{2}{\beta}+\lt(1-\frac{1}{\beta^{2}}\rt)\ln\lt(\frac{1+\beta}{1-\beta}\rt)\rt]\drm{t}^r
\\
&+\frac{1}{2}\int_{0}^{t}
\frac{qc}{\beta}\lt[\frac{2}{\beta}+\lt(1-\frac{1}{\beta^{2}}\rt)\ln\lt(\frac{1+\beta}{1-\beta}\rt)\rt]\drm{t}^r
\\
&{+\frac{1}{2} }c^{2}\int_{0}^{t} 
(t-t^{r})\lt(\frac{2}{\beta}+\lt(1-\frac{1}{\beta^{2}}\rt)\ln\lt(\frac{1+\beta}{1-\beta}\rt)\!\rt)\drm{t}^r
\\
&-\frac{1}{2}\int_{0}^{t}
(1-\beta^{2})qc\lt(\frac{2}{\beta^{2}}-\frac{1}{\beta^{3}}\ln \lt(\frac{1+\beta}{1-\beta}\rt)\!\rt)\drm{t}^r\Bigg].
\end{split}
\end{align}
{The first and third lines cancel manifestly.
 All terms but one simple integral cancel also when we add the second and fourth lines.
This yields
\begin{align}\label{eq:44}
F_{0}^{(4)}(t)=\frac{1}{4}\varkappa^4 e^{2}  \!\int_{0}^{t}\! q c\drm{t}^r.
\end{align}
}
 We have arrived at  (\ref{eq:F4}).

\end{appendix}

%%%%%%%%%%%%%%%%%%%%%%%%%%%%%%%%%%%%%%%%%%%%%%%%%%%%%%%%%%%%%%%%%%
%%%%%%%%%%%
%%%%%%%%%%% And finally: The bibliography 
%%%%%%%%%%%
%%%%%%%%%%%%%%%%%%%%%%%%%%%%%%%%%%%%%%%%%%%%%%%%%%%%%%%%%%%%%%%%%%


\newpage


\begin{thebibliography}{[9999999]}\footnotesize{

\bibitem[Bop1940]{BoppA}
\vskip-5pt
        Bopp, F.,
        \textit{Eine lineare Theorie des Elektrons},
        Annalen Phys. \textbf{430}, 345--384 (1940).
\bibitem[Bop1943]{BoppB}
\vskip-5pt
        Bopp, F.,
        \textit{Lineare Theorie des Elektrons. II}, Annalen Phys. \textbf{434}, 573--608 (1943).

\bibitem[Bor1933]{BornA} 
\vskip-7pt
	Born, M., 
	\textit{Modified field equations with a finite radius of the electron}, Nature \textbf{132}, 282 (1933).

\bibitem[Bor1934]{BornB} 
\vskip-7pt
	Born, M., 
	\textit{On the quantum theory of the electromagnetic field,} Proc. Roy. Soc. A \textbf{143}, 410--437 (1934).


\bibitem[BoIn1933]{BornInfeldA} 
\vskip-7pt
	Born, M., and Infeld, L.,
	\textit{Electromagnetic mass}, Nature \textbf{132}, 970 (1933).


\bibitem[BoIn1934a]{BornInfeldB} 
\vskip-7pt
	Born, M., and Infeld, L.,
	\textit{Foundation of the new field theory},
	Proc. Roy. Soc. London \textbf{A 144}, 425--451 (1934).

\bibitem[BoIn1934b]{BornInfeldCa} 
\vskip-7pt
	Born, M., and Infeld, L.,
	\textit{On the quantization of the new field equations. Part I}, 
Proc. Roy. Soc. London A \textbf{147}, 522--546 (1934).

\bibitem[BoIn1935]{BornInfeldCb} 
\vskip-7pt
	Born, M., and Infeld, L.,
	\textit{On the quantization of the new field equations. Part II},
 Proc. Roy. Soc. London A \textbf{150}, 141--166 (1935).

\bibitem[BKTZ2021]{BKTZ}\vskip-7pt
        Burtscher, A.,
	Kiessling, M.K.-H.,
       and
      Tahvildar-Zadeh, A. S.,
  \textit{Weak second Bianchi identity for static, spherically symmetric spacetimes with timelike singularity},
    Class. Quantum Grav. \textbf{38}, 185001(31pp) (2021).
\bibitem[CKP2019]{CKP}
\vskip-7pt
        Carley, H.K., Kiessling, M.K.-H., and Perlick, V.,
        \textit{On the Schr\"odinger spectrum of a Hydrogen atom with electrostatic Bopp--Land\'e--Thomas--Podolsky 
          interaction between electron and proton}, 
        Int. J. Mod. Phys. A \textbf{34}, 1950146 (23pp.) (2019).
\bibitem[CaKi2024]{CarKie}
\vskip-7pt
        Carley, H.K., and Kiessling, M.K.-H.,
        \textit{A study of the radiation-reaction on a point charge that moves along a constant applied electric field
        in an electromagnetic
        B{\tiny\sc{opp}}-L{\tiny\sc{and\'e}}-T{\tiny\sc{homas}}-P{\tiny\sc{odolsky}} vacuum},
        pp.323--336 in: ``Physics and the Nature of Reality: Essays in Memory of Detlef D\"urr,''
	A. Bassi et al., Ed.,  Springer, Heidelberg (2024).
\textbf{49},  445202 (19pp.) (2016).
\bibitem[Dir1938]{Dirac}
\vskip-5pt
	Dirac, P.A.M., 
        \textit{Classical theory of radiating electrons},
        Proc. Royal Soc. London \textbf{167}, 148--169 (1938).
\bibitem[Dys1952]{Dyson}
\vskip-5pt
Dyson, F., 
 \textit{Divergence of perturbation theory in quantum electrodynamics}, 
 Phys. Rev. \textbf{85}, 631--632 (1952).
\bibitem[GHW2009]{GHW} 
\vskip-.3truecm
        Gralla, S.E.,
        Harte, A.,
        and
        Wald, R.M.,
        \textit{A Rigorous Derivation of Electromagnetic Self-Force},
        Phys. Rev. D \textbf{80}:024031 (2009).

\bibitem[Fey1948]{Feynman}
\vskip-.3truecm
Feynman, R.P.,
\textit{Relativistic Cut-Off for Quantum Electrodynamics}, 
Phys. Rev. \textbf{74}, 1430--1438 (1948). 

\bibitem[GPT2015]{GratusETal}
\vskip-.3truecm
	Gratus, J., Perlick, V., and Tucker, R.W.,
        \textit{On the self-force in Bopp--Podolsky electrodynamics},
        J. Phys. A: Math. Theor. \textbf{48}, 435401 (28pp.) (2015).
\bibitem[Hetal2021]{VuMaria}
\vskip-.3truecm
	Hoang, V., Radosz, M., Harb, A., DeLeon, A.,  and Baza, A.,
        \textit{Radiation reaction in higher-order electrodynamics},
        J. Math. Phys. \textbf{62}, 072901 (31pp.) (2021).
\bibitem[Jac1975]{JacksonBOOKb} % 
\vskip-7pt
        Jackson, J.D.,   
                \textit{Classical electrodynamics}, 
       	J. Wiley \& Sons, New York 
	$2^{nd}$ ed. (1975).

\bibitem[Kie2012]{KieCMP}
\vskip-5pt
	Kiessling, M.K.-H.,
\textit{On the quasi-linear elliptic PDE
       $-\nabla\cdot(\nabla{u}/\sqrt{1-|\nabla{u}|^2}) = 4\pi\sum_k a_k \delta_{s_k}$ in physics and geometry},
     Commun. Math. Phys. \textbf{314}, 509--523 (2012);
        \textit{Correction} ibid. \textbf{364}, 825--833 (2018).

\bibitem[Kie2019]{KiePRD}
\vskip-5pt
	Kiessling, M.K.-H.,
        \textit{Force on a point charge source of the classical electromagnetic field},
        Phys. Rev. D \textbf{100}, 065012 (2019);
        \textit{Erratum} ibid. \textbf{101}, 109901(E) (2020).
\bibitem[KTZ2025]{KTZonBLTP}
\vskip-7pt
	Kiessling, M.K.-H.,
        and
        Tahvildar-Zadeh, A. S.,
        \textit{Bopp-Land\'e-Thomas-Podolsky electrodynamics as initial value problem},
        to be submitted (2026).

\bibitem[Lan1941]{Lande}
\vskip-5pt
	Land\'e, A., 
        \textit{Finite Self-Energies in Radiation Theory. Part I},
        Phys. Rev. \textbf{60}, 121--126 (1941).
\bibitem[LaTh1941]{LandeThomas}
\vskip-5pt
	Land\'e, A., and Thomas, L.H.,
        \textit{Finite Self-Energies in Radiation Theory. Part II},
        Phys. Rev. \textbf{60}, 514--523 (1941).

\bibitem[Laz2019]{Lazar}
\vskip-5pt
        Lazar, M., 
        \textit{Green functions and propagation in the Bopp--Podolsky electrodynamics},
        Wave Motion \textbf{91}, 102388 (2019).
\bibitem[LaLe2020]{LazarLeckA}
\vskip-5pt
        Lazar, M., and Leck, J.,
        \textit{Second gradient electromagnetostatics: A nonsingular electromagnetic field theory},
        Annals Phys. (NY) \textbf{423}, 168330 (2020). 
\bibitem[LaLe2020]{LazarLeckB}
\vskip-5pt
        Lazar, M., and Leck, J.,
        \textit{Second gradient electromagnetostatics: Electric point charge, electrostatic and magnetostatic dipoles},
        Symmetry \textbf{12}, 1104 (2020). 
\bibitem[Lor1904]{LorentzENCYCLOP}
\vskip-5pt
        Lorentz, H.A., 
                {\it Weiterbildung der Maxwell'schen Theorie: Elektronentheorie.},
        Encyklop\"adie d. Mathematischen Wissenschaften ${\bf V}2$,
        Art. 14, pp. 145--288 (1904).
\bibitem[Mil1998]{MillerBOOK} % 
\vskip-5pt
        Miller, A. I., % Arthur I.                                                                                      
                {\sl Albert Einstein's special theory of relativity},
        Springer, New York (1998).
\bibitem[PMD2006]{PMD}
\vskip-5pt
        de Parga, A., Mares, R., and Dominguez, S.,
        {\it An unphysical result for the Landau--Lifshitz equation of motion for a charged particle},
        Rev. Mex. Fis. {\bf 52}, 139--142 (2006).

\bibitem[PaVi1949]{PauliVillars}
\vskip-5pt
Pauli, W., and Villars, F., 
\textit{On the Invariant Regularization in Relativistic Quantum Theory}, Rev. Mod. Phys. \textbf{21}, 431--444
(1949). 
\bibitem[Pod1942]{Podolsky}
\vskip-5pt
        Podolsky, B.,
        \textit{A generalized electrodynamics. Part I: Non-quantum},
        Phys. Rev. \textbf{62}, 68--71 (1942).
\bibitem[PPV2011]{PoissonETal}
\vskip-5pt
   Poisson, E., Pound, A., and Vega, I.,
   \emph{The motion of point particles in curved spacetime},
 Living Rev. Rel. \textbf{14},7(190) (2011).
\bibitem[Sch1994]{QEDbook}
\vskip-7pt
Schweber, S.,
    \textit{QED and the man who made it: Dyson, Feynman, Schwinger, and Tomonaga}, 
    Princeton University Press (1994).

\bibitem[QuWa1997]{QuinnWald}
\vskip-7pt        
Quinn, T.C., and Wald, R.M.,
\textit{Axiomatic approach to electromagnetic and gravitational radiation reaction of particles in curved spacetime},
Phys. Rev. D \textbf{56}, 3381--3394 (1997).

\bibitem[Sch1935]{Erwin}
\vskip-7pt
        Schr\"odinger, E.,
        \textit{Contribution to Born's new theory of the electromagnetic field},
        Proc. Royal Soc. London A \textbf{150}, 465--477 (1935).

\bibitem[Spo2004]{Spohn}
\vskip-7pt
       Spohn, H.,
       \textit{Dynamics of charged particles and their radiation fields},
       Cambridge UP (2004).
\bibitem[Zay2014]{Zayats}
\vskip-5pt
        Zayats, A.E.,
        \textit{Self-interaction in the Bopp-Podolsky electrodynamics:
          Can the observable mass of a charged particle depend on its acceleration?},
        Annals Phys. (NY)\textbf{342}, 11--20 (2014).
}
\end{thebibliography}
\end{document}